\documentclass{jfm}
\usepackage{graphicx}
\usepackage{subcaption}
\usepackage{epstopdf, epsfig}
\usepackage{hyperref}
\usepackage{gensymb}
\usepackage{verbatim}
\usepackage{xcolor}
\usepackage{amsmath}
\usepackage{enumitem}

\usepackage{float}

\shorttitle{Shock, and microjet impact}
\shortauthor{R. Subramanian, Z. Yang, F. Roman{\`o}, and O. Coutier-Delgosha}

\title{Bubble collapse near a wall. \\ Part 1: An experimental study on the impact of shock waves and microjet \newline on the wall pressure}

\author{Roshan Kumar Subramanian\aff{1},
  Zhidian Yang\aff{2}, Francesco Roman{\`o}\aff{2},
 \and Olivier Coutier-Delgosha\aff{1,2}\corresp{\email{ocoutier@vt.edu}}}

\affiliation{\aff{1}Department of Aerospace, and Ocean Engineering, Virginia Polytechnic Institute and State University, Blacksburg, VA 24060, USA
\aff{2}Univ. Lille, CNRS, ONERA, Arts et Metiers Institute of Technology, Centrale Lille, UMR 9014 - LMFL - Laboratoire de Mécanique des Fluides de Lille - Kampé de Fériet, F-59000 Lille, France
}

\begin{document}

\maketitle

\begin{abstract}

In this first out of three parts, the pressure exerted by a cavitation bubble collapsing near a rigid wall is studied experimentally. A single cavitation bubble is laser generated by optical breakdown in a water basin. The bubble grows, subsequently collapses, thereafter it experiences a second growth, and finally a second collapse. The emission of shock waves and the liquid jet formation due to the non-spherical collapse occur during or at the end of each bubble collapse depending on the stand-off ratio $\gamma$, defined as the distance of the bubble centroid from the wall divided by the bubble equivalent radius. The phenomenological characterization of near-wall cavitation is investigated in detail, with a special focus on the emission of multiple shock waves from a single non-spherical collapse. We also detail the dominant shock mechanisms, either tip or torus collapse, for the several value of $\gamma$ considered in this paper. Our experiments use high-speed imaging techniques and Schlieren imaging to visualize the microjet interface, and the shock wave, respectively, for different stand-off ratios. On the one hand, the microjet evolution is described by tracking either the far-wall interface of the bubble or the near-wall vaporization emerging along the microjet path for large enough $\gamma$. On the other hand, the shock wave is captured as a composite image holding multiple shock positions for a single propagating shock wave occurring at the end of the collapse. A detailed quantitative analysis of the microjet interface and the shock wave velocities is reported in the paper, together with the corresponding impact time on the wall. The experimental investigation is further complemented by measuring the liquid pressure using a wall-mounted sensor and a needle hydrophone installed far from the wall. The impact times of the shock wave and the microjet observed with the high-speed visualizations are then compared with the peak pressure recorded by the wall sensor and the hydrophone to identify the dominant contribution upon a change of $\gamma$. The dominant contribution to pressure stresses on the wall is found to correlate with three collapse regimes we identify. This is expected to have important implications on the cavitation erosion mechanisms.

\end{abstract}


\section{Introduction}
 \vspace{5pt}
Cavitation is a widely studied subject in the fluid dynamics community due to the significant destructive effects caused in engineering and biological systems \citep{hydraulic,biological}. Cavitation is observed in various industries like space  \citep{Space}, ocean \citep{Ocean}, cleaning \citep{Cleaning}, and medical to name a few \citep{Medical}. We can observe cavitation in the turbo-pumps, and turbines of the rocket engines \citep{rocket_engine}, marine propellers of submarines \citep{marine_propellers}, pipe lines \citep{pipeline}, valves \citep{Valves}, bearings \citep{bearings}, combustion engines \citep{combustionengine}, and many other engineering or hydraulic systems in which cavitation is found to cause structural, and performance issues \citep{StrucDamage}. The damage caused by cavitation implies material deformation and, for the most catastrophic events, material loss leading to erosion \citep{StrucErosion}. In such a case, the erosion of a solid surface is due to the residual stress on the material originating from the grain boundaries. The slip lines eventually develop into cracks, thereby leading to mass loss \citep{Gao2019}. Several studies investigated erosion due to cavitation, however it is only recently that cavitation erosion is being harnessed for good namely for agricultural crop cleaning \citep{agriculture}, water treatment \citep{Cleaning}, kidney stones removal \citep{kidneystones}, and cancer tumor ablation \citep{Medical}.

Cavitation was first scientifically reported by Euler in 1754 during his study of turbo-machinery \citep{Euler}, and later described as a phenomenon of phase transition due to localized low pressure and/or high temperature \citep{Brennen}. The occurrence of cavitation is often observed in form of cloud of cavitation bubbles, which makes difficult to study the mechanisms at the origin of cavitation. Hence, theoretical researches focused on the simple case of a single cavitation bubble. Several attempts were made to model cavitation, including the pioneering study of \cite{Raleigh}, whose prediction of cavitation by high pressure pulses was based on the assumption of liquid being homogeneous, inviscid, and incompressible. Other initial efforts to model cavitation are due to \cite{Parsons1919, Kornfeld1944} who noted that the unstable surface cavities are developed contrasting the initial idea of cavities being either spherical or semi-spherical, and later \cite{Gilmore} and \cite{Hickling1964} theoretically included compressibility effects, surface tension, and viscosity contrast for the case of a single cavitation bubble collapse. The strong pressure pulse at the end of the bubble collapse leads to the rapid increase of pressure inside the bubble, which in turn leads to the emission of shock waves \citep{Tomita}. The exertion of strong pressure by the propagating spherical shock wave has been shown to be one of the mechanisms of erosion. \cite{Hickling1964} has shown that the cavitation damage, and the shock wave impact are highly associated. 

If the bubble is far from a boundary, the shock front is observed as spherical, while if the bubble collapses near a wall or a fluid-fluid interface, the spherical symmetry of the collapse is broken. As a result, non-spherical shock waves can be produced, as well as a liquid jet that pierces through the bubble interface leading to a toroidal collapse. Owing to its minute structure, \cite{Benjamin1966} proposed to refer to the liquid jet as `microjet', which nomenclature is widely employed in modern scientific literature. This is defined as the liquid structure typically funnel shaped, piercing the distal end of the bubble interface, and passing through the proximal end before hitting the nearby boundary \citep{Benjamin1966, Lindau2003, Arie1998}. Since the first observations of \cite{Parsons1919}, further confirmed and extended by \cite{Kornfeld1944}, such a microjet --- initially referred to as `hydraulic blows' --- is investigated to inquire if the jet mechanism plays an important role in damaging a solid surface near the bubble cavitation. The early experimental research on jets \citep{Naude, Benjamin1966, Tiru, Plesset} tend to state that the jet plays a significant role in the cavitation erosion, however a direct causation between localized damage and the emergence of a microjet has never been rigorously proven by experimental measurements. 

To understand which mechanism among the shock or the jet impact dominates the wall erosion process, a clear experimental quantitative characterization of these fluid dynamic phenomena is targeted in our study. The pressure magnitude will be recorded and associated with the source producing it. Moreover, Schlieren photography is the photo-optical technique predominantly used to visualize the shock waves and will be hereinafter widely employed to characterize the dynamics of shock fronts. The first ever visualization of shock waves from collapsing cavitation bubbles was done by \cite{Mundry}. After them, numerous studies on the visualization of shocks from collapsing cavitation bubbles were published \citep{Tomita, Vogel1988, Supponen2017Shock}, however, to the best of the authors' knowledge, there is no study that unequivocally measures the impact time of the shock on the rigid wall, and compares it with the jet impact and the pressure associated with both phenomena.

There are different methods to create cavitation in a lab environment, however the creation of individual cavitation bubbles is required to understand the mechanisms of cavitation erosion. Four different methods of single cavity generation technique are commonly employed: (i) tube-arrest method \citep{Chesterman1952}, (ii) acoustic-generated method \citep{AcousticExp}, (iii) spark-induced method \citep{SPARKExp}, and (iv) laser-induced bubble generation method \citep{LASERExp}. Among these approaches, laser-induced bubble generation technique is used extensively in the experimental investigation of cavitation bubbles, and was first introduced by Lauterborn. A very repeatable, and precise bubble generations is possible using this technique, and that makes this a preferred choice among the cavitation researchers.

\cite{HemantRecent} investigated single bubble dynamics to understand cavitation damage, concluding that jets and the second (rebound) collapse are significant contributors. Recently, \cite{DularRecent} conducted experimental research on single cavitation bubbles, finding that jets play a significant role when 
$\gamma <$ 0.2, while the influence of microjets diminishes beyond 0.5 $<\gamma$. However, their study did not unequivocally quantified the effects of shock waves, leaving the dominant erosion mechanism unanswered. This research gap is being addressed in the current experimental paper.

Experimental investigations on cavitation bubbles alone may not provide a complete understanding of this physical phenomenon due to limitations in observing dynamic processes over different scales of time and length. These constraints are effectively addressed by numerical simulations, which are essential for comprehending the intricate nature of cavitation. Numerical simulations of single bubble dynamics are typically categorized as two-fluid models and single-fluid models, further classified into interface-tracking and interface-capturing methods \citep{Comp_Review}. Interface-capturing methods, such as the 'Level-Set Method' (LSM), and the 'Volume of Fluid' (VOF) methods excel in capturing the interface topology due to complex phenomena like the formation of reentrant jets during bubble collapse near surfaces. Many researchers in the past carried away the work on single bubble dynamics using these methodologies \citep{LSComp_3,LSComp_5,VOFComp_4}. Recently, the Level-Set Method (LSM) has gained widespread adoption in studies due to its ability to precisely depict evolving interfaces, which is essential for comprehending cavitation dynamics. In a notable example, \cite{Comp_1} employed LSM to simulate bubble topology at smaller stand-off distances, delineating various jet regimes. Moreover, \cite{Comp_2} focused on simulating laser-induced bubble collapse dynamics. Both VOF and LSM methods will be used in our follow-up papers (part 2, and part 3 respectively).

In the following section, \S\ref{sec:SetUp}, we will detail the experimental investigation set-up to capture and quantitatively characterize the emergence of shock waves and the near-wall collapse microjet that could both lead to cavitation erosion. The bubble evolution, the sources and symmetries of the shock waves, as well as the topology of the microjet interface are among the major focuses of \S\ref{sec:Results}. In \S\ref{sec:Dominant}, we will identify which of these two phenomena exerts the peak pressure on the nearby solid boundary for different bubble stand-off ratios $\gamma$ reproducing each experiment several times to associate a converged error bar to each experimental measurement in order to allow for robust cause-effect considerations. Finally, in \S\ref{sec:Conclusions}, a summary of our investigation and the corresponding conclusions will be presented.

\section{Experimental set-up and methods}\label{sec:SetUp}
\vspace{5pt}

\begin{figure}
  \centerline{\includegraphics[height=0.425\textwidth]{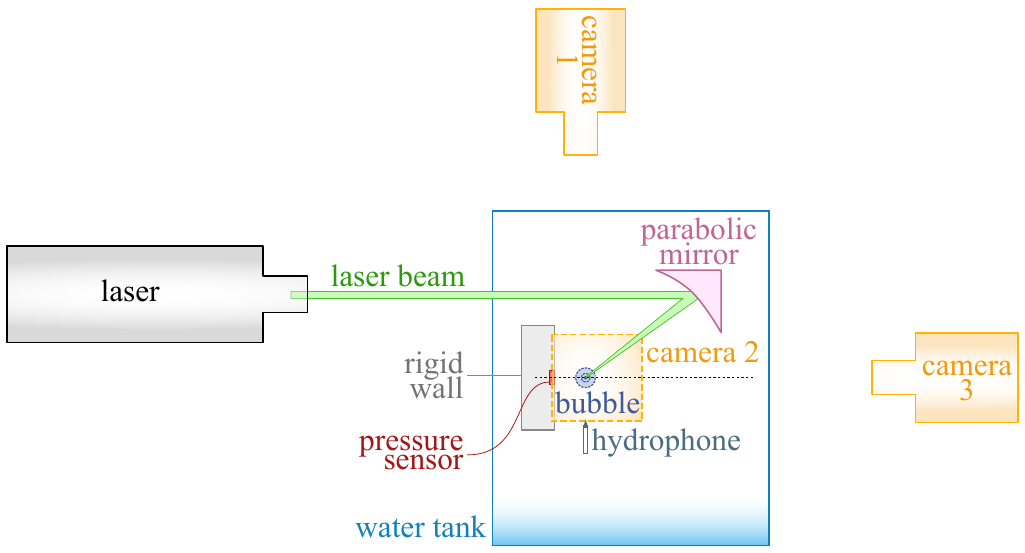}}
  \caption{Experimental schematic of laser-induced bubble generation.}
\label{fig:ExperimentalSetup}
\end{figure}
All experiments were performed in the `Cavitation, Propulsion, and Multiphase flow lab' at VirginiaTech, Blacksburg, VA (USA). Figure \ref{fig:ExperimentalSetup} shows the schematic of our laser-induced experimental setup used in this cavitation bubble study. This is also explained in our previous publication \citep{Ben}. We employ a 532 nm Q-switched Nd:YAG LASERs (Big Skyer LASER PN00148300) to create the cavitation bubble. The laser source is connected to the beam expanders (Thorlabs BE05-532) employed before focusing the beam by means of a $60\degree$ off-axis parabolic mirror (Edmund Optics 35-554) placed inside the water tank. The distance between the beam expander (not shown in the schematics), and the center point of the parabolic mirror is $\sim$ 9.25 inches, while the distance between the beam expander and the front face of the water tank through which the laser beam enters the tank as shown in fig.\ \ref{fig:ExperimentalSetup} is $\sim$ 4 inches. The dimension of the glass water tank used in the experiment is 20.3$\times$20.3$\times$20.3 cm$\textsuperscript{3}$. The thickness of the glass in the water tank is 4.8 mm. We use three different visualizations namely top, side and front view by camera 1 (Phantom VEO 710), camera 2 (Phantom TMX 6410) and camera 3 (and Phantom VEO 440), respectively. The cameras are used with the 100 mm lens (Tokina AT-X M100 PRO D Macro), attached with the notch filter. The notch filters (Edmund Optics 86-130) are used to block the laser wavelength to protect the cameras from damaging, and to acquire better recordings. The microjet visualizations are best captures by camera 1, while shock waves are best visualized through camera 2. Whenever toroidal collapses occur, camera 3 becomes essential to visualize them, owing to the quasi-axi-symmetry of the bubble collapse with respect to the dashed black line in fig.\ \ref{fig:ExperimentalSetup}. The field of views used for the jet and shock visualizations are 11$\times$11 mm$\textsuperscript{2}$, and 15$\times$15 mm$\textsuperscript{2}$ respectively. The rigid wall used in this experiment is a clear acrylic block of thickness 12.1 mm, length 8.5 mm, and width 7.5 mm. The rigid wall near which the laser focusing produces a cavitation bubble is controlled by using a translation stage and a micrometer (Thorlabs XRN25P-K1) which enables to precisely adjust the stand-off distance between the bubble centroid and the wall surface. The experiments are performed using the de-ionized water.

\begin{figure}
\centering
  \includegraphics[height=0.575\textwidth]{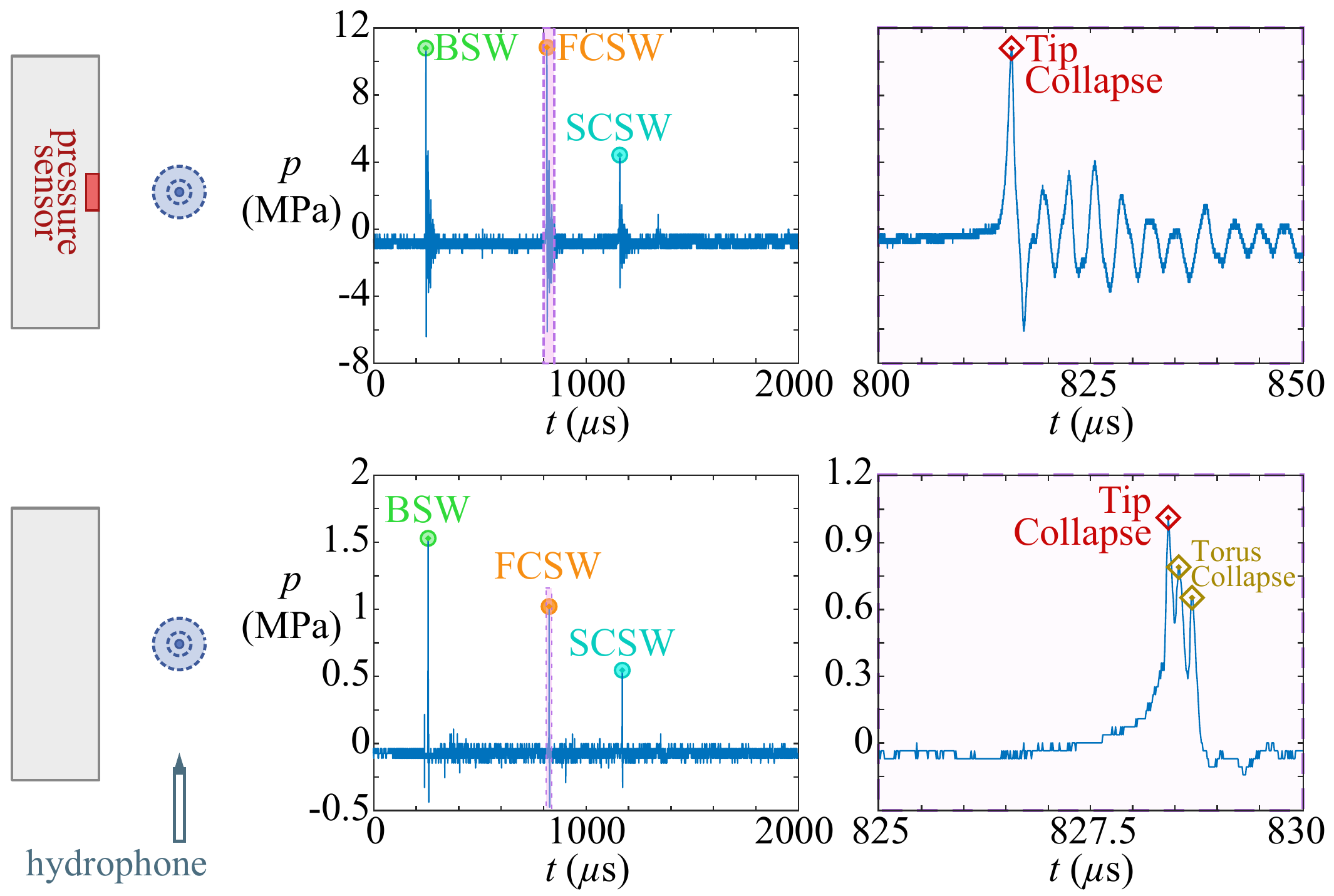}
  \caption{Example of pressure signals of the wall sensor (top) and the hydrophone (bottom) for $\gamma=2.19$.}
\label{fig:ExamplePressures}
\vspace{20pt}

  \subfloat[$\gamma = 2.19$]{\vspace{-0.2cm}\includegraphics[height=0.25\textwidth]{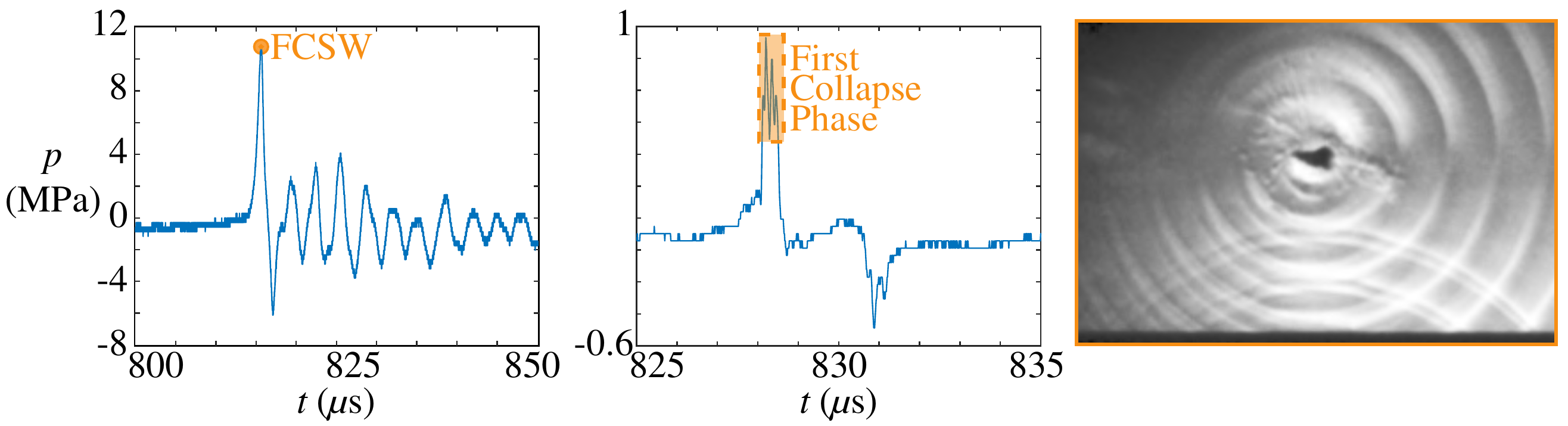}} \\
  \subfloat[$\gamma = 1.60$]{\vspace{-0.2cm}\includegraphics[height=0.25\textwidth]{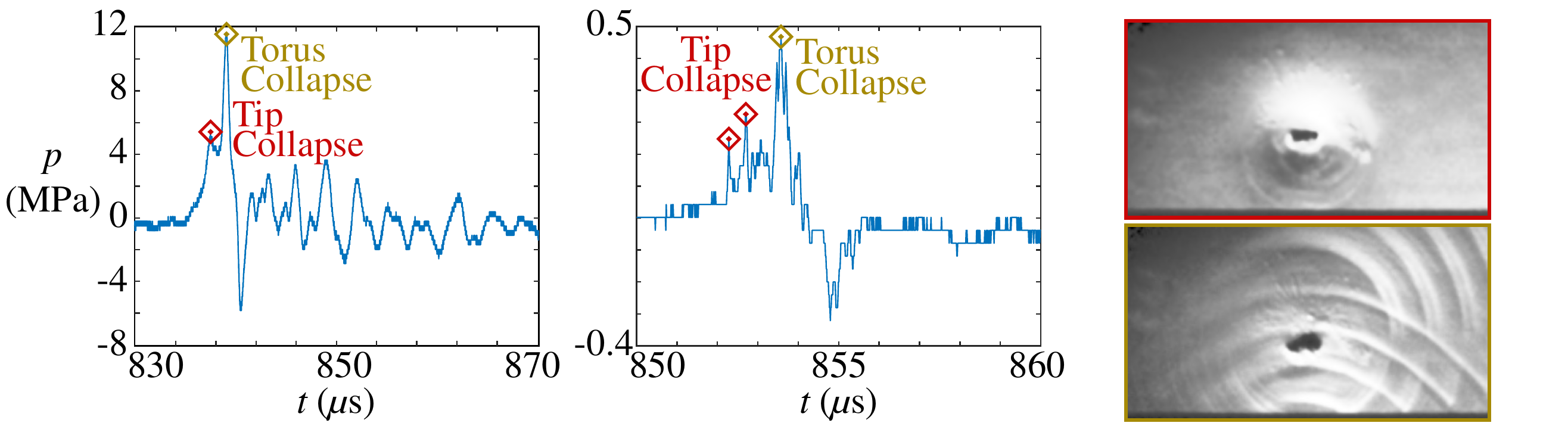}} 
\caption{Examples of pressure signals of the wall sensor (left), hydrophone (middle), and corresponding Schlieren visualizations (right) for (a) $\gamma$=2.19, and (b) $\gamma$=1.60 respectively. Mixed tip-and-torus collapse mechanisms emerge in the intermediate $\gamma$.}
\label{fig:FirstCollapseShocks}
\end{figure}

The magnitudes of the pressure forces are measured using two sensing devices, a pressure sensor flush-mounted on the wall, and a hydrophone located at 45 mm from the bubble collapse point (see fig.\ \ref{fig:ExperimentalSetup}). The pressure sensor that is used to quantify the pressure magnitude on the wall employs PCB piezotronics, and the model number is W105C22 with a sensitivity of 137.2 mV/MPa. The sensor was calibrated specifically for this experiment from the manufacturer to ensure the highest accuracy. The precision hydrophone (model number NH0200 0.2 mm needle) was used to measure the acoustic transients. The Nd-YAG LASER sources, and the cameras shown in fig.\ \ref{fig:ExperimentalSetup} are controlled and triggered by a pulse generator (BNC Model 577), with the laser source and the cameras triggered at the same time. The time for the laser source emitting the beam on receiving a trigger, and getting focused to create optical breakdown in water using the parabolic mirror is $\sim$ 130 $\mu$s. All the experimental pressure data are collected using the oscilloscope (BNC Model P2025) that can be saved using the flash drive for further processing. The sampling rate of the oscilloscope to record the signal is set at the maximum level of 500 Mega Samples per second (MSa/s). The oscilloscope is triggered when it receives the signal above the threshold value due to the pressure rise. The threshold value is set to a very small value of $\sim$ 20 mV to trigger the oscilloscope, and it gets triggered from the shock wave emitted during the bubble generation. This is usually referred to as \textit{breakdown shock wave} (BSW). The trigger point is set a little away $\sim$ 250 $\mu$s from the beginning of the oscilloscope signal to ensure to record the complete signals including BSW. This can be seen in fig.\ \ref{fig:ExamplePressures} in which the BSW is not at the beginning of the recorded oscilloscope signal in both sensor, and hydrophone signals. Figure \ref{fig:ExamplePressures} shows the typical signal recorded by the wall sensor, and the hydrophone. The first and second collapses induce a \textit{first collapse shock wave} (FCSW) and a \textit{second collapse shock wave} (SCSW) exerting pressure on the wall by the emission of such shock waves. As it will be detailed in the following sections, the bubble collapse near a wall can involve a tip and a torus collapse. We anticipate, however, that the tip collapse occurs from the trapped vapor pocket between the distal, and proximal interface (closer to the rigid wall) of the bubble while the jet is piercing the bubble. Both of them can occur during the collapse of a cavitation bubble near the rigid wall and they give rise to two shock waves. Since these shock mechanisms occur in a very short interval of time ($< 2 \ \mu$s), the pressure sensor on the wall may not be able to distinguish among them for large $\gamma$. The hydrophone is therefore used to identify the different shock mechanisms as shown in the 'zoom-in' image of the hydrophone signal in fig.\ \ref{fig:ExamplePressures}. More details about the interpretation and understanding of such signals are given in the results section.

A qualitative difference is however observed when considering the shocks at small and large stand-off ratios, as demonstrated in fig.\ \ref{fig:FirstCollapseShocks}. The occurrence and interactions of the tip and torus collapses emerging during the bubble lifetime are discussed and rationalized in the results section.

\subsection{Bubble interface visualization}
\vspace{5pt}
\begin{figure}
  \centering
  \subfloat[$\gamma = 0.78$]{\includegraphics[height=0.225\textwidth]{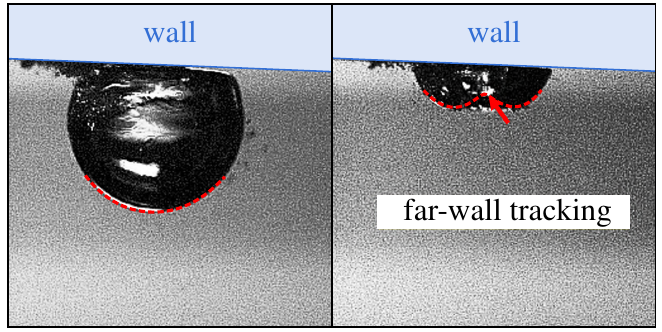}} \quad \subfloat[$\gamma = 2.01$]{\includegraphics[height=0.225\textwidth]{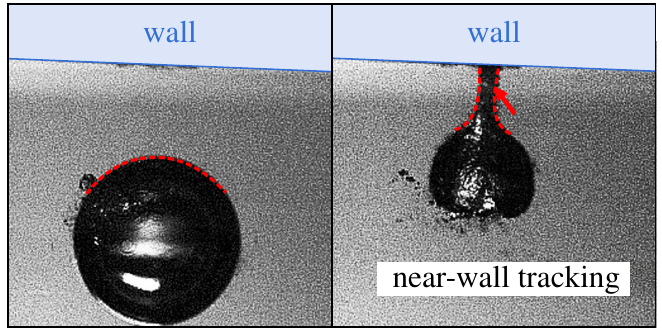}}
  \caption{Interface high-speed visualizations. The left panels in (a) and (b)depict the cavitation bubble at its maximum radius while the right panels show a snapshot during the first collapse phase. The far- and near-wall interfaces tracked by high-speed visualizations are highlighted by the red dashed lines and pointed at by the red arrows. (a) Far-wall interface tracking for $\gamma = 0.78$. (b) Near-wall interface tracking for $\gamma = 2.01$.}
\label{fig:MicroJet}
\end{figure}

The liquid jet piercing through the vapour bubble cannot be directly quantified by means of the visualization techniques employed in this study. For this reason, rather than measuring the liquid microjet velocity, we estimate it by employing the velocity of the bubble interface. A detailed numerical study that shows how the liquid microjet velocity correlates with the bubble interfacial velocity will be reported in Part 2 and Part 3 of this series of papers. 

For $\gamma \leq 1.24$, the dynamics of the bubble growth will bring the bubble very close to the wall. Hence, during the first collapse phase, its interface will mainly deform far from the wall (see fig.\ \ref{fig:MicroJet}(a)). In such cases, it will be possible to measure the velocity of the far-wall interface deformation as an estimate of the microjet velocity. On the other hand, for $\gamma > 1.41$, the bubble will start collapsing far enough from the wall to develop a jet around which vaporization occurs (see fig.\ \ref{fig:MicroJet}(b)). The corresponding liquid microjet velocity will therefore be estimated by tracking the near-wall interface for $\gamma > 1.41$. The bubble interface visualizations are captured at 77k\,fps with an exposure time of 12.99\,$\mu$s.

\subsection{Stand-off ratio $\gamma$}
\vspace{5pt}
\begin{figure}
\centering
  \subfloat[experiment]{\includegraphics[height=0.29\textwidth]{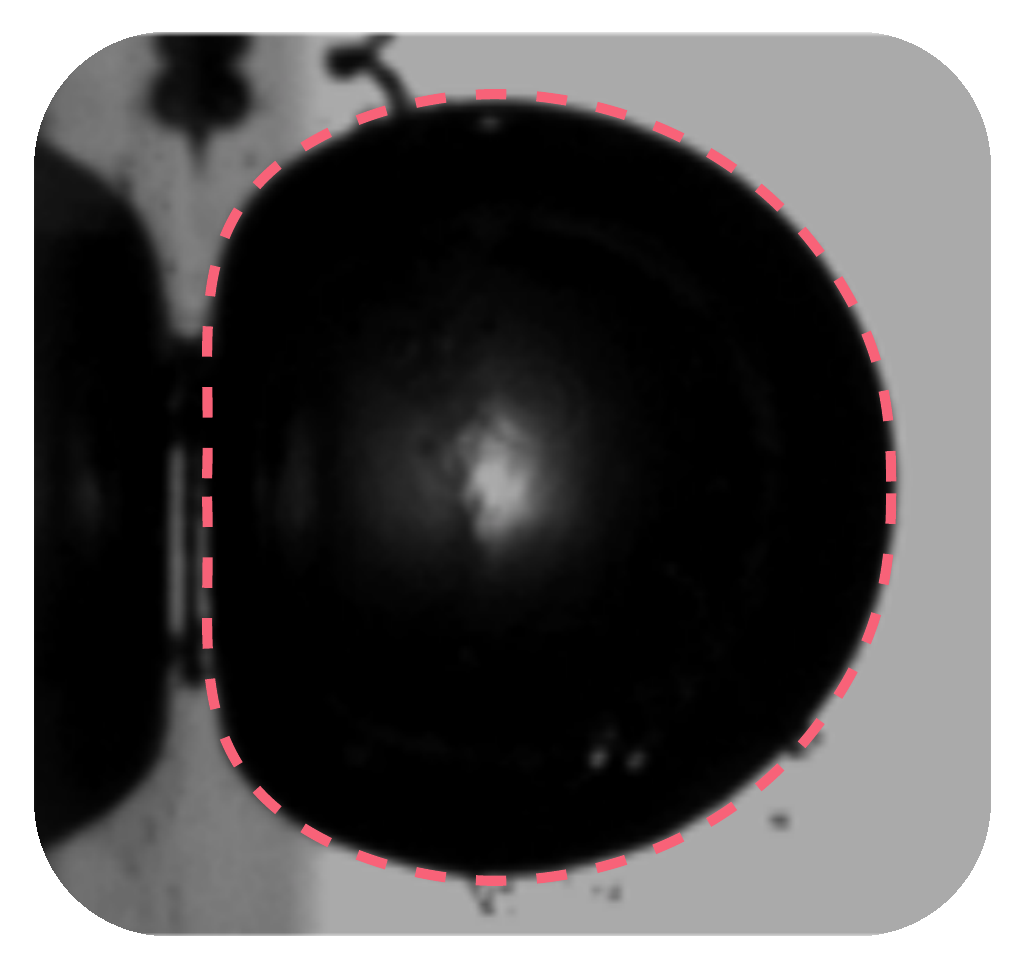}} \quad
  \subfloat[fitting]{\includegraphics[height=0.29\textwidth]{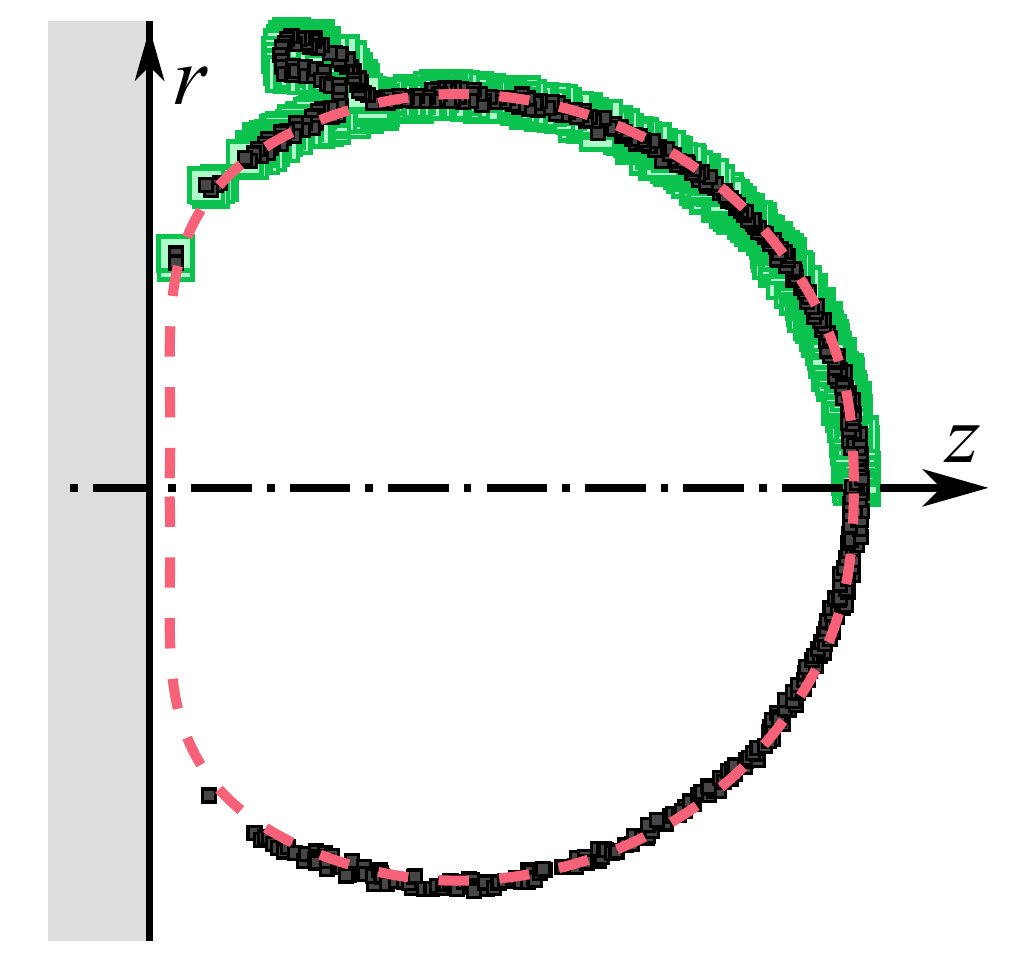}} \quad
  \subfloat[$L(t)$ and $R_{eq}(t)$]{\includegraphics[height=0.31\textwidth]{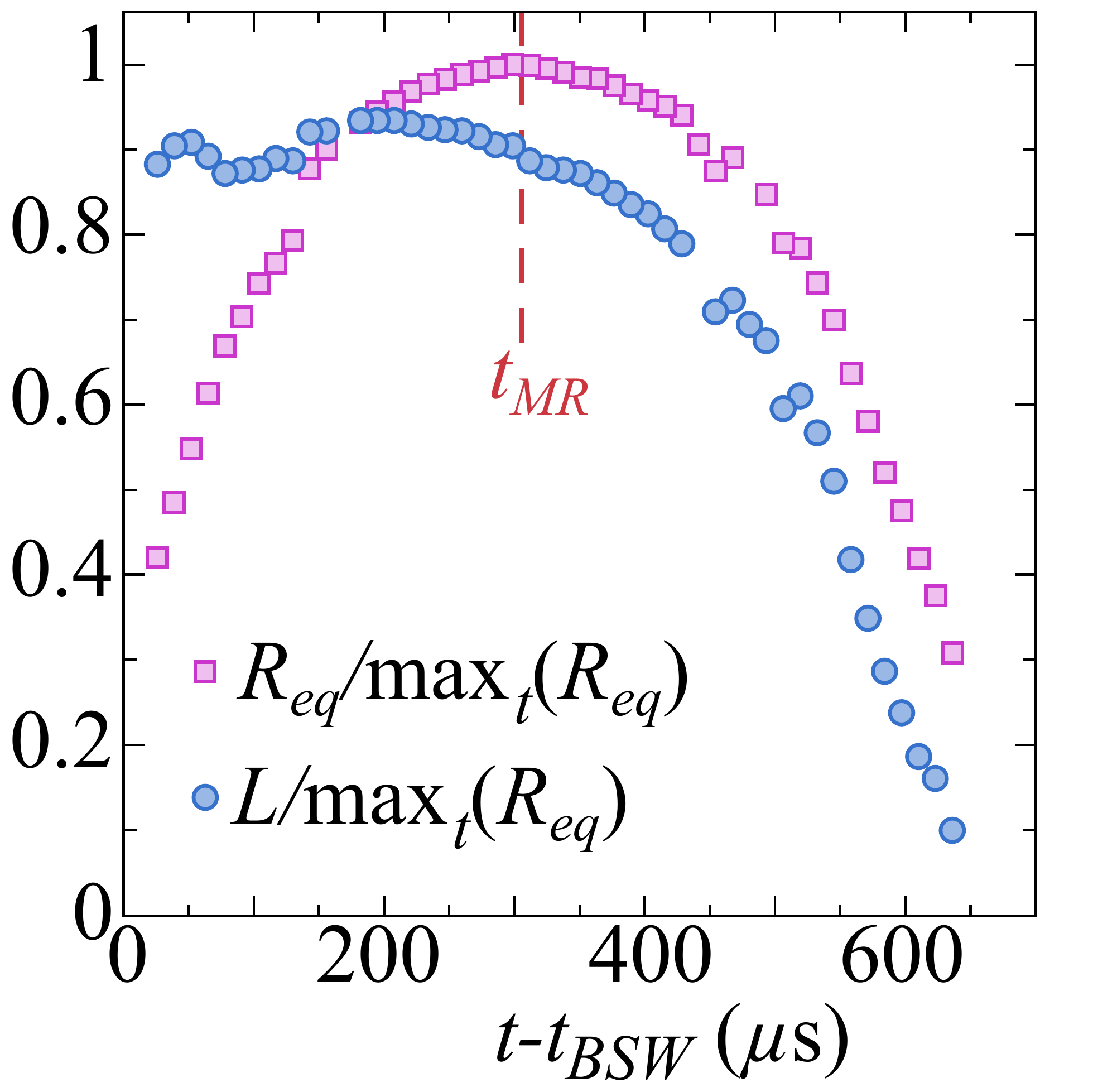}}
\caption{(a) Experimental top view of a cavitation bubble for $\gamma = 0.91$. (b) Experimental interface location (black markers), set of mirrored interfacial coordinates (green markers), fitting function (red dashed) by employing \eqref{eq:fitting}, and wall location (gray). (c) Evolution of $L$ and $R_{eq}$ with time. The reference equivalent radius is $\max_{t}[{R_{eq}}(\gamma=0.91)] = 2.74$\,mm.}
\label{fig:FittingGamma}
\end{figure}
The stand-off ratio $\gamma$ is assumed in our study as the control parameter to vary for investigating the bubble cavitation near a wall. Such a non-dimensional parameter is defined as the stand-off distance $L$, i.e. the distance between the bubble centroid and the wall, divided by the bubble equivalent radius $R_{eq}$, i.e. the radius of the equivalent sphere having the same volume of the cavitation bubble. As $L$ and $R_{eq}$ evolve over time due to the multiple bubble growths and collapses, $\gamma$ is uniquely defined as
\begin{equation}
    \gamma = \frac{L\left(t_{MR}\right)}{\max\limits_{t} \left(R_{eq}\right)}, \qquad \forall t \in [t_{BSW},\ t_{FCSW}],
\end{equation}
where $t_{MR}$ denotes the time at which the vapour bubble attained its maximum radius after its generation at $t_{BSW}$ and before the first collapse at $t_{FCSW}$, i.e. $t_{MR} = t({R_{eq}=\max_{t} (R_{eq})})$. The stand-off ratio is calculated by determining $L(t)$ and $R_{eq}(t)$ from the experimental interface visualization $\forall t \in [t_{BSW},\ t_{FCSW}]$. The following algorithm is used to compute $\gamma$: \vspace{1cm}
\begin{enumerate}[labelwidth=0.75cm,leftmargin=1cm,itemindent=0cm,align=right]
    \item[I.\ \ ] The video of the bubble growth from the top view (camera 1, see fig.\ \ref{fig:ExperimentalSetup}) is cropped to the range $t \in [t_{BSW},\ t_{FCSW}]$ and all corresponding frames are extracted. An example is shown in fig.\ \ref{fig:FittingGamma}(a).\vspace{0.1cm}
    \item[II.\ \ ] The bubble's interface's coordinates $(r_b,z_b)$ are extracted for each frame by exploiting the sharp interfacial identification due to the index refraction contrast between liquid water and water vapour. The corresponding result is demonstrated by black markers in fig.\ \ref{fig:FittingGamma}(b).\vspace{0.1cm}
    \item[III.\ \ ] The problem is simplified by assuming that the bubble growth is axi-symmetric, which allows us to reflect part of the interfacial coordinates $(r_b,z_b)$ with respect to the axis, as shown by the green markers in fig.\ \ref{fig:FittingGamma}(b).\vspace{0.1cm}
    \item[IV.\ \ ] An in-house MATLAB code developed to fit the bubble's interface is implemented for employing the following fitting function
    \begin{equation}\label{eq:fitting}
        r^2 = a_0^2 - \left(z-a_1\right)^2 \times \left(1 + \frac{a_2}{z^{a_3}}\right),
    \end{equation}
    where $a_0$, $a_1$, $a_2$, and $a_3$ are four constants to fit at each frame, $z$ and $r$ denote the axial and radial coordinates of the fitted interface, assuming that the bubble grows axi-symmetrically until $t_{MR}$. The result of fitting the blue markers is demonstrated by the red dashed line in fig.\ \ref{fig:FittingGamma}(b).\vspace{0.1cm}
    \item[V.\ \ ] The volume and the axial location of the bubble centroid are computed for the axi-symmetric fitting function \eqref{eq:fitting}, and their evolution in time is demonstrated in fig.\ \ref{fig:FittingGamma}(c).\vspace{0.1cm}
    \item[VI.\ \ ] The stand-off ratio is finally computed by identifying the time for which the equivalent radius reaches its maximum, i.e. $t_{MR}$, and evaluating the ratio between $L\left(t_{MR}\right)$ and $\max_{t} (R_{eq})$. \vspace{0.2cm}
\end{enumerate}

Since the parabolic mirror is fixed in position, the bubble is generated every time at the same location within the water tank. As the laser power is kept constant, the bubbles produced for smaller $\gamma$ do not experience a significant shrinking of their maximum equivalent radius. The maximum bubble radius measured by our fitting algorithm ranges from $\max_{t}[{R_{eq}}(\gamma=0.65)] = 2.66$\,mm to $\max_{t}[{R_{eq}}(\gamma=2.19)] = 2.77$\,mm. The value of $\gamma$ is controlled by displacing the wall, hence by adjusting $L$. It results that all our experiments fall within the range $\gamma \in [0.65,2.19]$.

\subsection{Schlieren setup and shock wave tracking}
\vspace{5pt}
\begin{figure}
  \centerline{\includegraphics[height=0.235\textwidth]{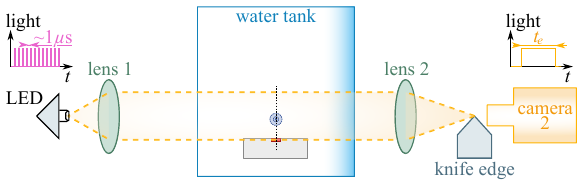}}
  \caption{Simple Schlieren setup used for visualizing shock waves. The light pulse of the LED is depicted in magenta and it corresponds to $\sim 1$\,$\mu$m, while the exposure time $t_e$ of the camera is depicted in orange.}
\label{fig:SimpleSchlieren}
\end{figure}
The shock waves have higher velocity in water than air, which makes them challenging to be visualized in water. Moreover, the refractivity is higher for water due to higher density than air. After considering several potential techniques for visualizing the shock waves, Schlieren photography has been chosen in this study owing to the sharpness of the achieved visualizations thanks to the sensitivity of the Schlieren technique \citep{Schlieren}. The Schlieren tenchnique detects the density gradients in the transparent media, in our case de-ionized water. 

Various configurations have been proposed in the literature for Schlieren imaging by setting up different arrangements for the lenses, mirrors, and the light sources. In this study, we use the so-called `simple Schlieren setup' depicted in fig.\ \ref{fig:SimpleSchlieren}. The light source used in this setup is a LED pulsed driver from 'LightSpeed Technologies, model number: HPLS-36DD18B'. The two lenses (lens 1 and 2 in fig.\ \ref{fig:SimpleSchlieren}) are achromatic doublets from Thor Labs with model numbers AC508-075-A (F=75 mm) and ACT508-300-A (F=300 mm), respectively. A knife-edge is used as the explicit cut-off tool of the refracted light beam due to shock wave in the water before being focused on the image plane of the high-speed camera.

\begin{figure}
  \centerline{\includegraphics[height=0.4\textwidth]{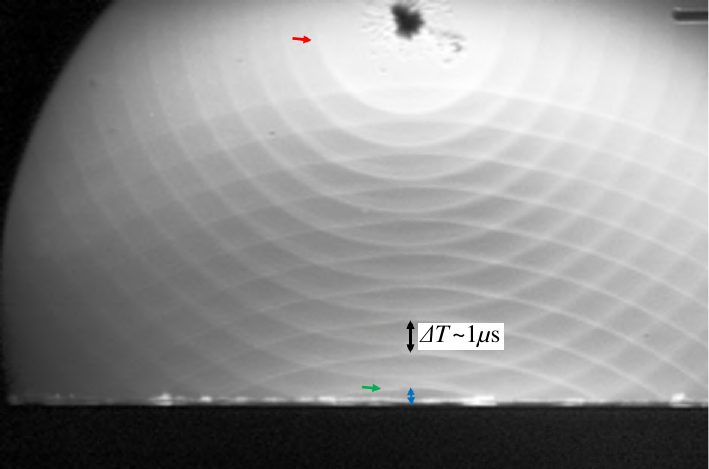}}
  \caption{Schlieren visualization of a quasi-spherical shock wave as a composite image captured using microsecond light flashes for $\gamma = 4$. The used exposure time is 19\,$\mu$s. Red arrow indicates the propagating single spherical shock wave. Green arrow points to reflected single spherical shock wave from the wall. Black double arrow indicates the time interval $\Delta T \sim 1$\,$\mu$m between the two fronts. Blue double arrow shows the space that is extrapolated to capture the shock impact time.}
\label{fig:SphericalShock}
\end{figure}

In order to capture experimentally the time at which the shock waves will impact on the wall, high-speed visualizations are combined taking advantage of multiple shock visualizations with a frequency given by the led pulse. One flash per microsecond is the frequency $f_{LED}$ used to capture the shock at different instants of time along its propagation. The multiple shock locations are then combined in a single photo whose exposure time is greater than a the flash pulse. An example of quasi-spherical shock wave generated for $\gamma = 4$ is depicted in fig.\ \ref{fig:SphericalShock}, showing an example of the composite image with the different shock locations. Figure \ref{fig:SphericalShock} is captured with the exposure time of $T_{e} \sim$ 19 $\mu$s,  hence we can see the $N_{front} = T_{e} \times f_{LED} = 19$ shock locations, including the propagating and the reflected shock wave positions. This demonstrates how clearly we can estimate the time of impact of the shock wave on the rigid boundary by means of our experimental setup. The red arrow in fig.\ \ref{fig:SphericalShock} points at the first propagating shock wave position, and the green arrow points at the first reflected shock wave position of the same spherical shock wave emitted from the first collapse. By identification of the different shock positions and knowing $\Delta T=f_{LED}^{-1}=T_{e}/N_{front}$, the shock speed can therefore be estimated counting how many wave fronts $N_{impact}$ are needed for the impact. In case of the fig.\ \ref{fig:SphericalShock}, $N_{impact}\in [9,\ 10]$. The blue double-arrow identifies the region where the impact of the shock wave on the wall occurs for the first time and the exact time of impact is here approximated by extrapolation of the previous shock fronts. Knowing that the distance between the first shock front and the wall is $L_{s-w}= 0.01285\  m$, the shock propagation velocity in fig.\ \ref{fig:SphericalShock} is $U_{shock} = L_{s-w}/(N_{impact}\times\Delta T) = (1486.48 \pm 36.44) \ m/s$.

The frame rates used to capture the shock waves range from 150k\,fps to 180k\,fps with an exposure time of 4\,$\mu$s to 6\,$\mu$s for $\gamma$ = 0.65 to 2.19. The frame rates are increased and the exposure times are reduced for adjusting the accuracy of our measurements to the lower stand-off ratios. We always make sure that the captured shock images will hold between 4 to 6 shock fronts in a single composite image. This is considered sufficient to grant the necessary accuracy for drawing a corresponding conclusion. Moreover, to ensure repeatability, each experiment is repeated 10 different times for each $\gamma$, giving rise to a normal distribution of our data. All the respective error bars needed to support our claims are calculated with a 95\% confidence interval and reported for each set of experiments upon a change of $\gamma$.

\section{Results and discussion}\label{sec:Results}
\vspace{5pt}
\subsection{Cavitation bubble evolution near a rigid wall}
\begin{figure}
  \centerline{\includegraphics[height = 0.275\textwidth]{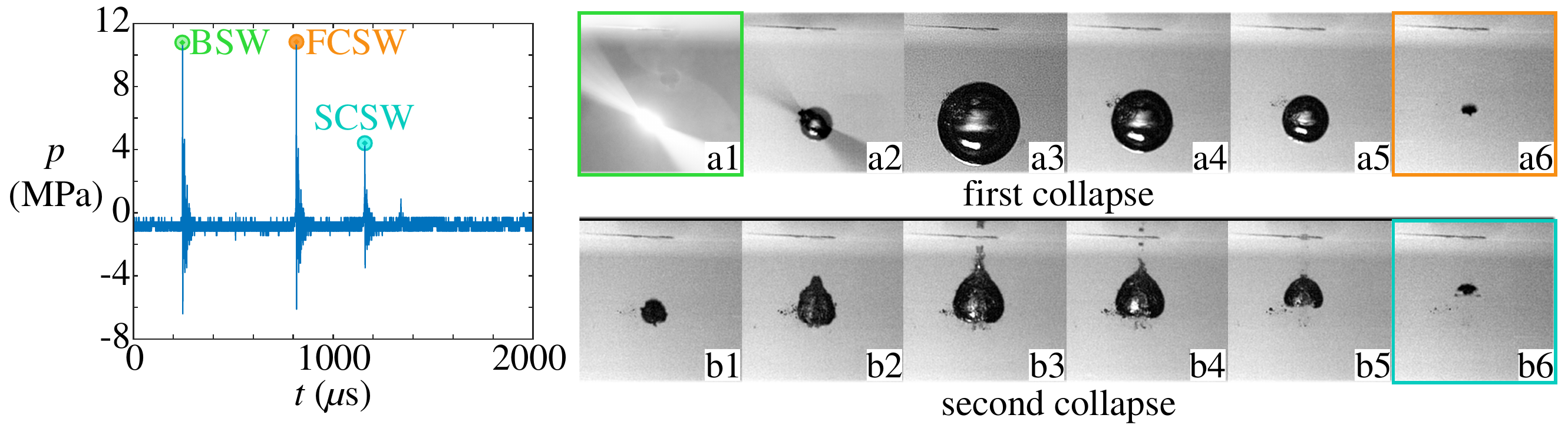}}
  \caption{Cavitation bubble evolution near a rigid wall for $\gamma$=2.19}
\label{fig:CavitationBubbleGamma2d19}
\end{figure}

\vspace{5pt}
The cavitation bubble is generated by the optical breakdown of the liquid using the laser. The laser getting focused for generating the bubble forms the focusing cone in the liquid after being reflected by the parabolic mirror as shown in fig.\ \ref{fig:ExperimentalSetup}. At the focal point, a high-pressure (see green marker in the left panel of fig.\ \ref{fig:CavitationBubbleGamma2d19}) and high-temperature plasma gets formed \citep{Plasma-temp}. Its initial evolution significantly depends on the laser energy being focused, hence by the intensity of the laser pulse, the focusing angle induced by the parabolic mirror, and the thermo-physical properties of the liquid. In our experiments, the expanded laser beam hitting the off-axis 60$\degree$ parabolic mirror helps to focus the laser in an almost elliptical structure, and further away from the glass tank's faces. As a result, after the initial phase transition occurred, the resulting vapour bubble growth is significantly impacted by this first formation phase \citep{Plasma-Shape}. A dedicated numerical study aimed at investigating how the initial conditions for the vapour bubble evolution influence its growth and collapse will be presented in Part 3 of this series of papers. Special attention to the non-linear absorption of the laser pulse by the liquid water will be paid in the numerical investigations of our subsequent manuscript, as we anticipate that this is a key element to correctly predict the formation and evolution of a vapour bubble.

The mechanical relaxation after the formation of the plasma leads to the formation of the shock wave at bubble generation known as \textit{breakdown shock wave} (BSW, see green box in fig.\ \ref{fig:CavitationBubbleGamma2d19}, \cite{BSW}). This initial shock wave is followed by the cavitation bubble growth shown in the right panels of fig.\ \ref{fig:CavitationBubbleGamma2d19}. Eventually, the cavitation bubble will collapse owing to the pressure inside the bubble being significantly lower than the ambient pressure in the liquid. A corresponding analysis of the influence of the vapour-to-liquid pressure ratio will be presented in Part 2, however we anticipate that, according to our simulations, the order of magnitude of the vapour-to-liquid pressure ratio expected for the current experiments $\approx 1/50$ when the vapour bubble attains its maximum equivalent radius.

When the bubble reaches its maximum size, the vapour pressure is at its minimum. At this stage, the BSW already propagated through the liquid, got reflected by the wall and the water pressure surrounding the vapour bubble returned approximately to the atmospheric pressure under which the vapour bubble has been generated. The \textit{first collapse} of the water bubble is therefore driven by the pressure difference between the inside and the outside of the bubble, featuring an initially slow evolution of the bubble interface (see `first collapse' in the right panel of fig.\ \ref{fig:CavitationBubbleGamma2d19}). The final phase of the first bubble collapse is associated to a very quick interfacial dynamics that leads to the formation of one or multiple shock waves (FCSW, see orange marker in the left panel of fig.\ \ref{fig:CavitationBubbleGamma2d19}). A visualization of the bubble right prior to collapse is highlighted by the orange box in fig.\ \ref{fig:CavitationBubbleGamma2d19}) and occurs at 569.42 $\mu$s from bubble generation time ($t_\text{BSW}=246.25 \ \mu s$).

The spherical symmetry, proper of bubbles formed in an unbounded medium, is broken by the presence of the wall. A corresponding pressure gradient is resulting in the liquid phase to compensate for the no-penetration at the wall. As a direct consequence, the bubble cavitation must collapse non-spherically due to the aforementioned pressure gradient across the bubble. Hence, the microjet visualized in the previous section occurs, significantly influencing the dynamics of the bubble collapse. The intricate interaction between bubble collapse, the corresponding shocks and the microjets are discussed in details in the following sections, as their relative occurrence allows the identification of the bubble collapse regimes of interest for our study. For $\gamma = 2.19$, the microjet interface is best observed during the second collapse (see figs.\ \ref{fig:CavitationBubbleGamma2d19}(b2--b4)).

After the first collapse, the vapour bubble undergoes a rebound that makes it growing up to almost its largest size. Owing to the presence of the wall, which induces a non-spherical collapse, the bubble is now toroidal in shape as the microjet pierced it, hence its rebound is strongly non-spherical (see cyan box in fig.\ \ref{fig:CavitationBubbleGamma2d19}). It follows a \textit{second collapse} phase leading to the production of shock waves (SCSW, see cyan marker in fig.\ \ref{fig:CavitationBubbleGamma2d19}). For $\gamma = 2.19$, the second collapse occurs at 912.53 $\mu$s from bubble generation ($t_\text{BSW}=246.25 \ \mu s$). A detailed analysis of first and second collapse, as well as of their corresponding regimes, will be presented in the following sections.

\subsection{Maximum pressure exerted on the wall}
 \vspace{5pt}

Among the main targets of our study there is the characterization and understanding of the maximum pressure exerted during the first and second collapse on the wall. Two main contributions have been identified and qualitatively discussed so far, namely: the shock waves and the microjet. The shock waves have been studied for decades in the literature, including theoretical predictions of their propagation \citep{Cole, ShockTheory2, ShockTheory1}. More recently, several investigations focused on the formation of the microjet during the non-spherical collapse near a wall and identified it as a potential source of strong pressure raise at the wall. The microjet is in fact predicted to travel at a velocity of the order of hundreds of meters per second. The numerical prediction done by \cite{Plesset} has shown that the microjet velocity can go up to $\sim$ 130-170 m/s, which is also in agreement with our numerical simulations (see Part 2 and 3), as well as with several other literature sources \citep{JetVel1, JetVel2}. This velocity magnitude in the microjet is speculated to produce a significant increase of pressure on the wall, even if this is still a controversial point in the scientific literature \citep{MJorShock1, MJorShock2, DularRecent}, especially when the aim is to identify which of the two mentioned mechanisms leads to the maximum wall pressure.
 
\begin{figure}
  \centerline{\includegraphics[height = 0.45\textwidth]{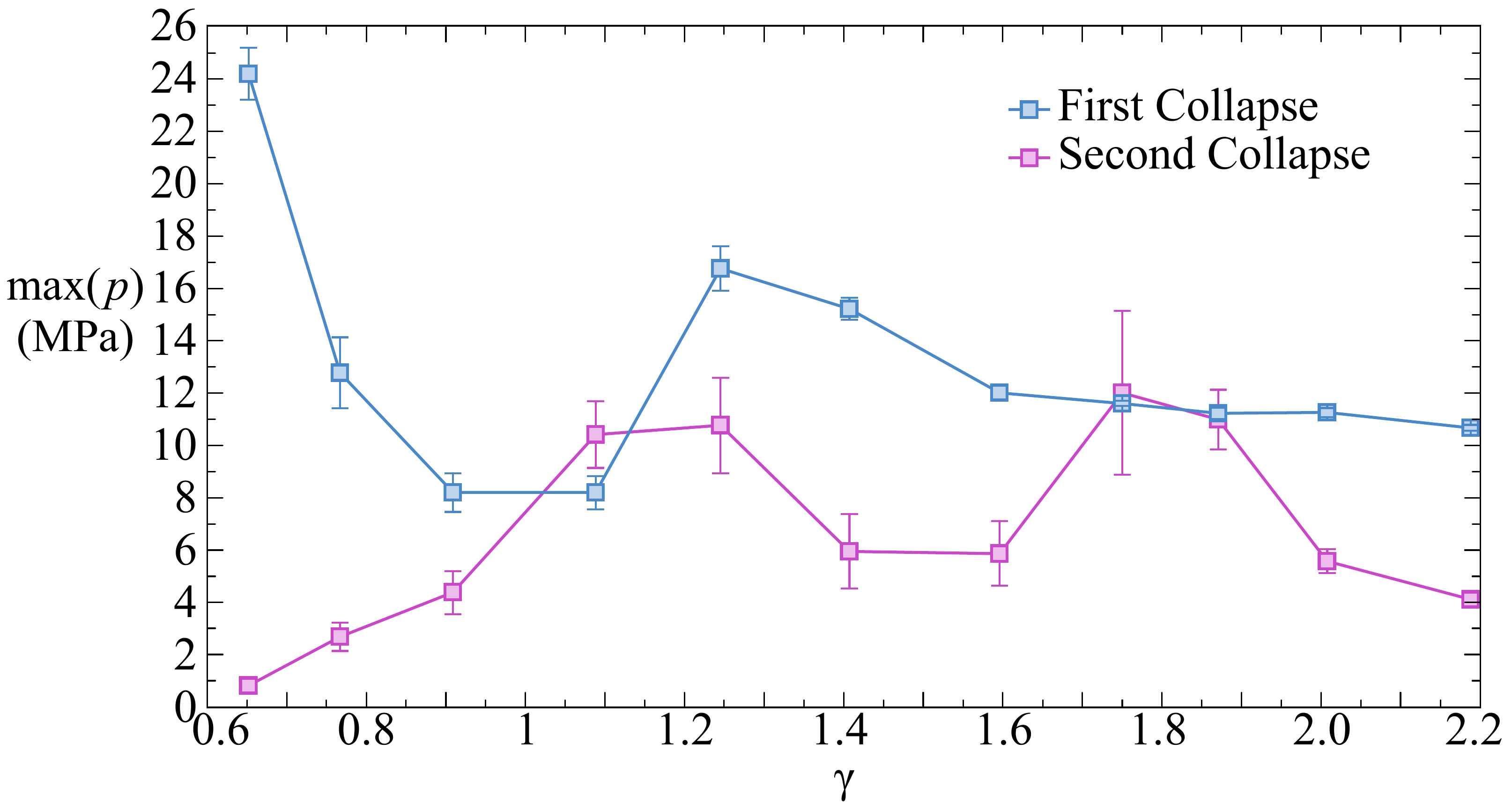}}
  \caption{Maximum pressure amplitude exerted by the first and second collapse of a single cavitation bubble on the wall for different stand-off ratios.}
\label{fig:WallPressure}
\end{figure}

This first part of our research will address such a question experimentally for a single bubble, identifying the regimes and scaling observed for the first and second bubble collapse. The maximum wall pressure exerted by the first (light blue), and the second (magenta) collapse are plotted in fig.\ \ref{fig:WallPressure}. The stand-off ratio is varied from $\gamma$=0.65 to 2.19. Among all the maximum wall pressures, the minimum value at the first collapse occurs at intermediate stand-off ratios $\gamma=\mathcal{O}(1)$, i.e. for $\gamma\in[0.91,\ 1.09]$ with $\min_{\gamma}[\max_t(p)] \approx 8.2$\,MPa. On the other hand, the maximum pressure occurs when the bubble collapses the closest to the wall, i.e. for $\gamma$=0.65 with $\min_{\gamma}[\max_t(p)] \approx 24.2$\,MPa. 

The maximum wall pressure trend with the stand-off ratio is interestingly non-monotonic as shown in fig.\ \ref{fig:WallPressure}. A maximum wall pressure rise is observed when the bubble collapses closer to the wall. Interestingly, the trend of increasing pressure gets hindered, and the maximum wall pressure declines reaching a minimum for $\gamma = \mathcal{O}(1)$. Thereafter, a sharp maximum pressure increase is observed upon an increase of $\gamma$, followed by a smoother decrease of $\max(p)$. Finally, fig.\ \ref{fig:WallPressure} demonstrates that the maximum wall pressure reaches a plateau for $\gamma \gtrapprox 1.75$.

The maximum pressure exerted on the wall by the second collapse is shown in fig.\ \ref{fig:WallPressure} in magenta color. An almost linearly-increasing trend is observed for the small stand-off ratios upon an increase of $\gamma$. Two local peaks are then observed for $\gamma \in [1.09,\ 1.24]$ and $\gamma \in [1.75,\ 1.87]$. It is important to note that the second collapse occurs from the compression of the rebounded bubble after the microjet pierces the bubble interface on both ends. This points out that a potential regime identification for the first collapse is expected to imply a correspondent set of regimes for the second collapse owing to the topological continuity in the bubble evolution for a given $\gamma$.

Considering the intricate dynamics and trends observed for the bubble lifetime and maximum wall pressure, a detailed quantitative analysis of the experimental measurements is required in order to relate the occurrence of the microjet, the evolution of the bubble interface and the corresponding shock waves during the first and second collapses. 

\subsection{Identification of relevant time scales}\label{sec:Dominant}
\begin{figure}
  \centerline{\includegraphics[height = 0.6125\textwidth]{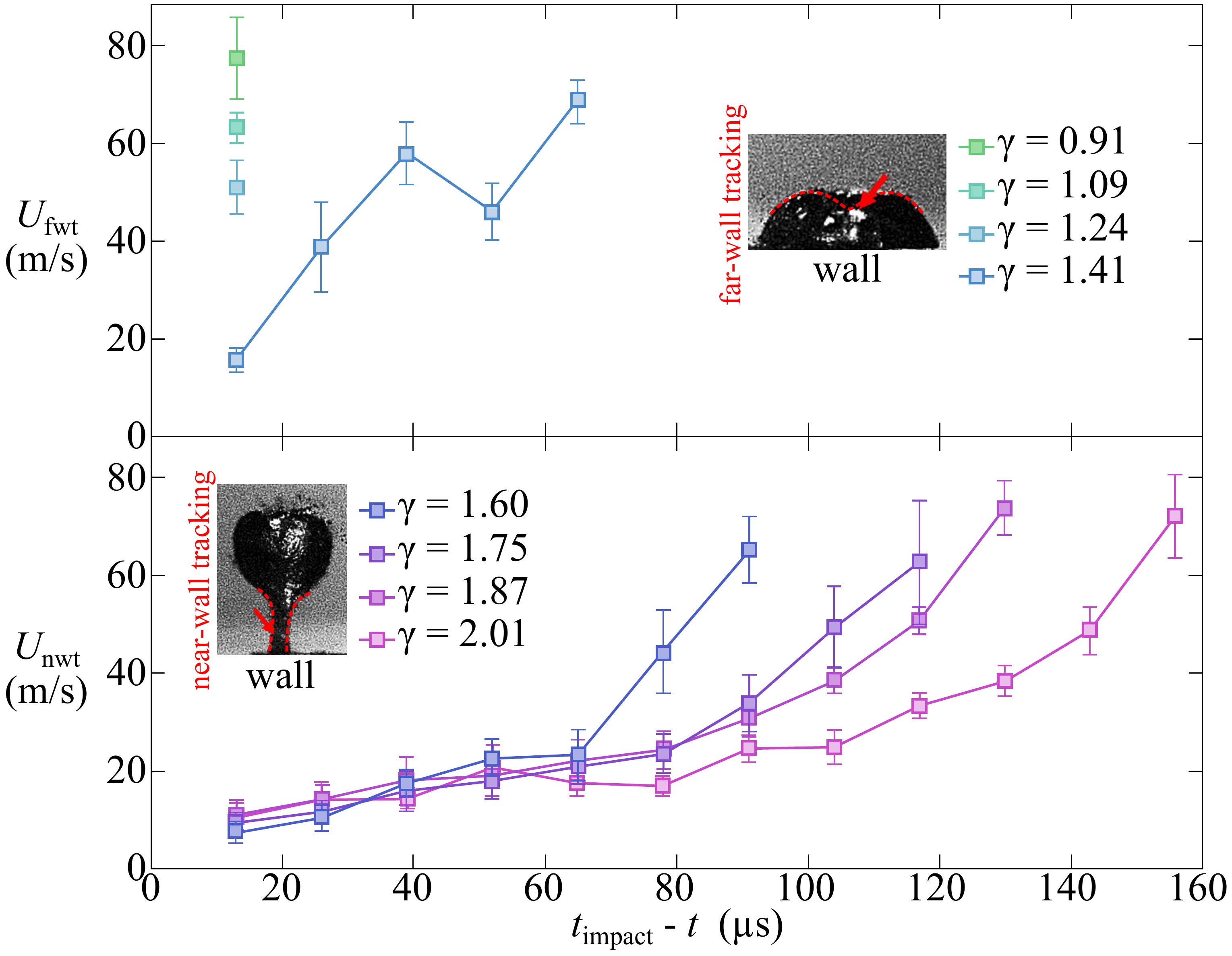}}
  \caption{Microjet interface velocity prior to impact on the wall. The top panel depicts the microjet interface estimated by means of the interfacial far-wall tracking (subscript `fwt'), while the bottom panel refers to the near-wall tracking (subscript `nwt'). }
\label{fig:JetInterface}
\end{figure}

\vspace{5pt}
The experimental investigation on the source of maximum wall pressure and the understanding of the observed trend with $\gamma$ passes through the accurate estimate of the corresponding time scales. The impact time of the microjet on the wall has been estimated by measuring the bubble interface velocity, assuming that this is representative of the fluid jet velocity and computing the corresponding time of impact for the microjet. Figure \ref{fig:JetInterface} depicts the evolution in time of the far- ($U_\text{fwt}$, top panel) and near-wall-interface ($U_\text{nwt}$, bottom panel) velocities measured with respect to the microjet impact time $t_\text{impact} - t$ upon a variation of $\gamma$. Characteristic velocities of the interface span over 10 to 80\,m/s with a decreasing trend upon approaching the wall (i.e. for $t_\text{impact} - t\rightarrow 0$). The initial deviations between various $\gamma$ for $U_\text{nwt}$ become negligible for $t_\text{impact} - t < 60\,\mu$s, signifying that the dynamics right prior bubble collapse is controlled by the near-wall dynamics and can be generalized regardless of the stand-off ratio. 

\begin{figure}
  \centerline{\includegraphics[height = 0.45\textwidth]{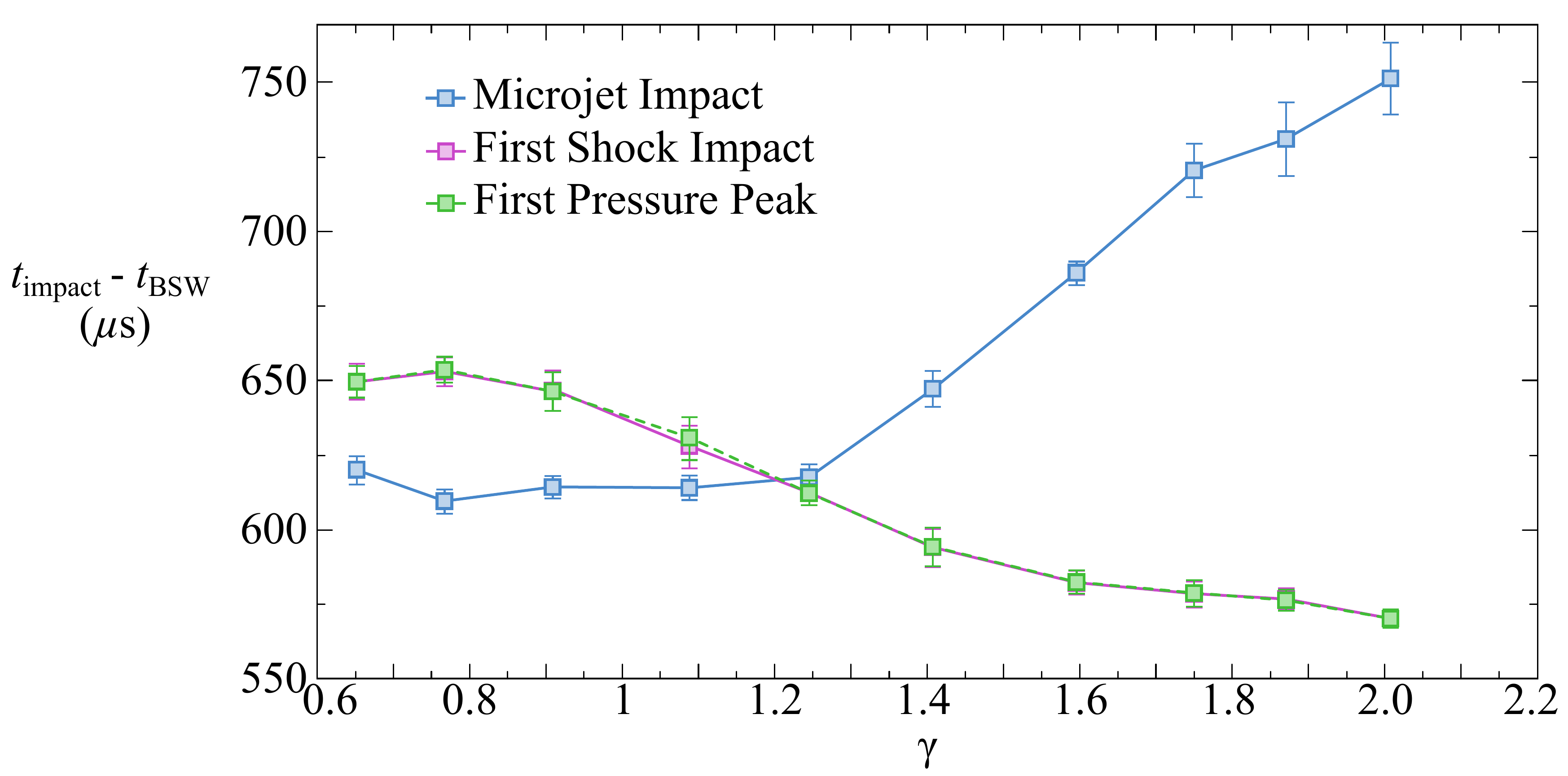}}
  \caption{Time of the jet and the first shock wave wall impact compared with the pressure peak time measured at the wall. All times are shifted by the bubble generation time ($t_\text{BSW}$).}
\label{fig:ImpactTime}
\end{figure}

The corresponding microjet impact time is depicted by light-blue squares in fig.\ \ref{fig:ImpactTime} and it increases with the bubble distance from the wall. This is well understood considering that the microjet impact strongly depends on the interface velocity and the position of the location of the bubble upon emerging of the microjet ultimately piercing it. Hence, the further the bubble gets pierced by the microjet, the longer the microjet will take to impact on the wall. The microjet impact is only weakly controlled by the interface deformation time scale and the corresponding transport velocity $\approx U_\text{fwt}, U_\text{nwt}$, which significantly decreases upon approaching the wall.

On the other hand, the shock wave impact time, measured for each stand-off ratio by identification of the shock front (see fig.\ \ref{fig:SphericalShock}), increases with $\gamma$ (see magenta markers in fig.\ \ref{fig:ImpactTime}). This trend cannot be easily understood by simple considerations of the relative position of the bubble with the wall, as it is controlled by an intricate dynamics of the bubble collapse. The characterization of such a dynamics and identification of first-collapse regimes will be discussed in the following section. We anticipate, however, that the topological conditions of the interface under which the shock-wave emerges play a crucial role in determining the bubble dynamics, hence the impact time of the shock on the wall. 

Comparing the impact time of microjet interface (light-blue markers), shock wave (magenta markers), and the occurrence of the pressure peak at the wall (green markers), fig.\ \ref{fig:ImpactTime} clearly demonstrates that the effect directly responsible for the pressure peak is the impact of the first-collapse shock wave on the wall. The almost identical time between shock-wave impact on the wall and maximum wall pressure peak is also observed for the second collapse (not shown). However, the role of the microjet in the dynamics of the bubble collapse is everything but negligible, and we will demonstrate that the largest value of the pressure at the wall strongly depends on the characteristic time scale at which the microjet pierces the bubble.

\subsection{Wall pressure and corresponding regimes for the first bubble collapse}
\vspace{5pt}

When collapsing in the absence of any pressure gradients, the bubble will collapse spherically. This produces a single outward propagating spherical shock wave at the end of the collapse. On the other hand, owing to a pressure gradient, the bubble collapses non-spherically. The presence of a rigid wall induces a pressure gradient that results in the non-spherical collapse leading to the emission of multiple propagating spherical shock waves at different instants of time \citep{Lindau2003, Brujan-Mechanism, Silvestre-Mechanism}. This was more recently studied by \cite{Supponen2017Shock}, who categorized the shock wave sources into \textit{jet collapse}, \textit{tip collapse}, and \textit{torus collapse}. 

Jet collapse is obtained as a jet impacting the near-wall interface of the bubble induces a shock due to the water-hammer pressure, and not due to the compression of trapped vapour pocket. The water-hammer pressure $p_h$ due to the jet impact can be defined as $p_h = \frac{1}{2}\rho c U_{jet}$, where $\rho$ is the liquid density, $c$ the celerity of sound in the liquid, and $U_{jet}$ the liquid microjet velocity \citep{WaterHammer}. This type of collapse occurs when the bubble is farther away from the boundary ($\gamma > 3.23$), and we do not observe it in our experiment due to the moderate $\gamma$ investigated in this study. 

Tip collapse is due to the small vapor pocket getting trapped at the proximal end of the collapse before the jet impact. On the other hand, torus collapse occurs when a toroidal bubble gets formed because of the piercing of its interface by the microjet and then collapses. According to \cite{Supponen2017Shock}, tip collapses occur for moderate stand-off ratios ($3.23 > \gamma > 1.14$), while toroidal collapse dominates for $\gamma < 1.14$.

\begin{figure}
\centering
  \subfloat[$\gamma = 0.65$]{\vspace{-0.2cm}\includegraphics[height=0.25\textwidth]{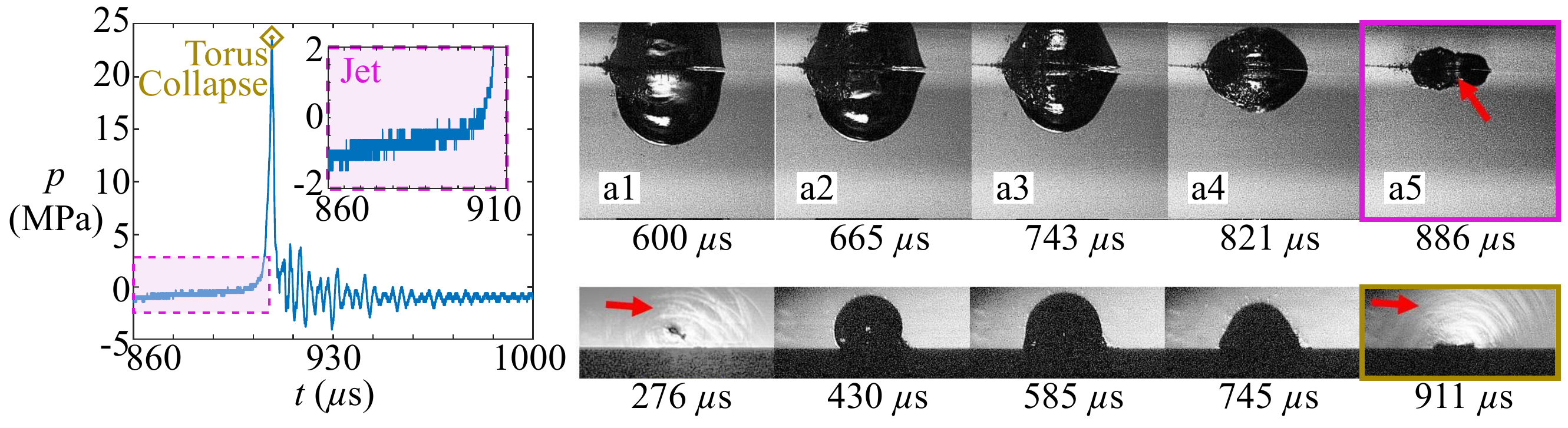}} \\
  \subfloat[$\gamma = 0.77$]{\vspace{-0.2cm}\includegraphics[height=0.25\textwidth]{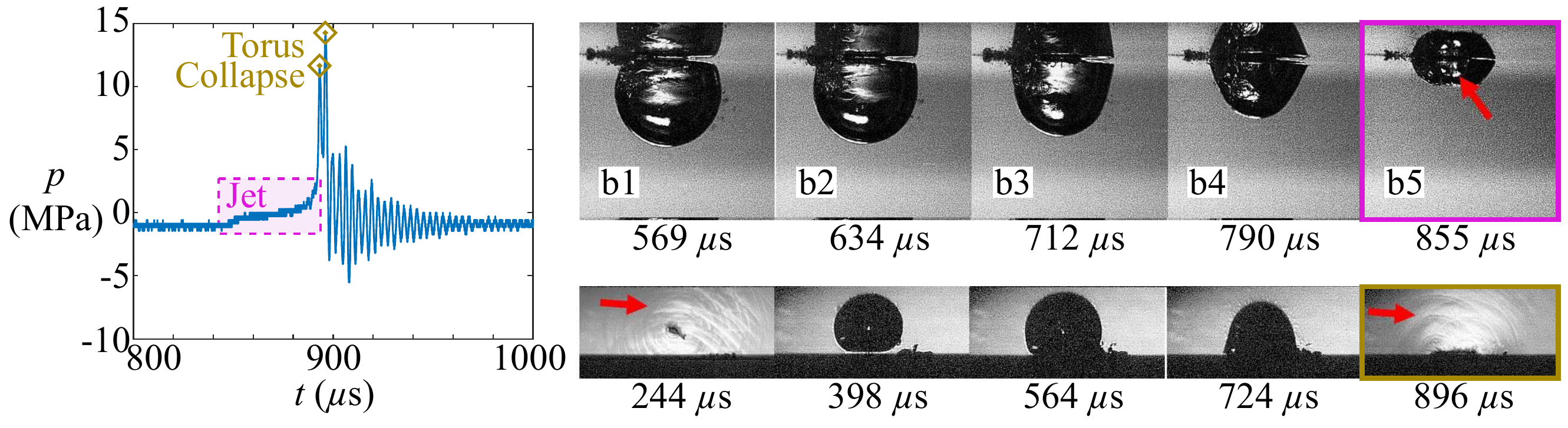}} \\
  \subfloat[$\gamma = 0.91$]{\vspace{-0.2cm}\includegraphics[height=0.25\textwidth]{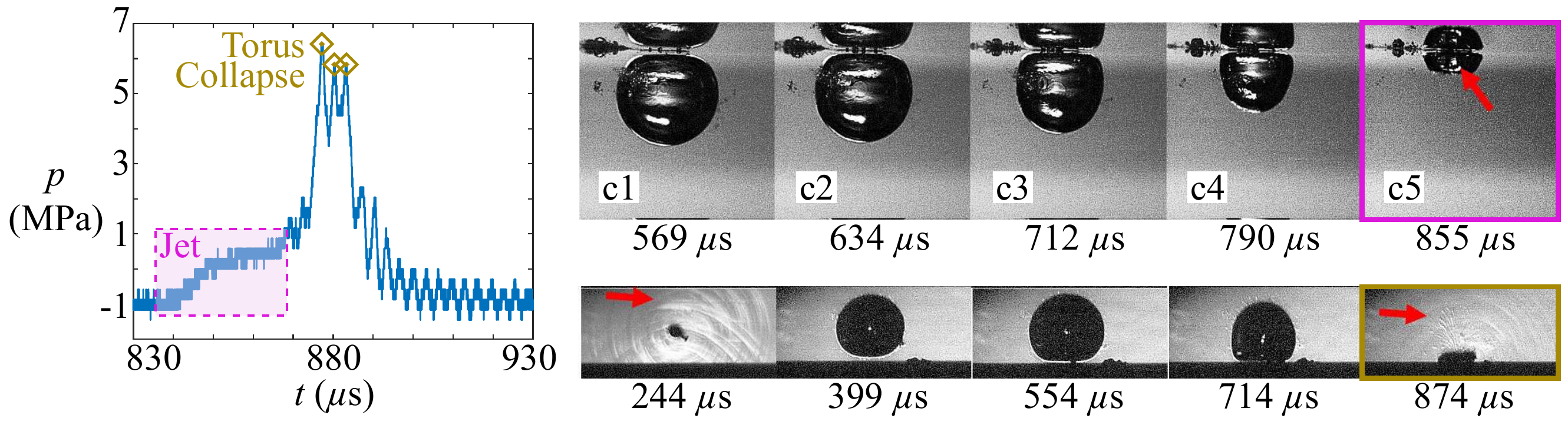}} \\
  \subfloat[$\gamma = 1.09$]{\vspace{-0.2cm}\includegraphics[height=0.25\textwidth]{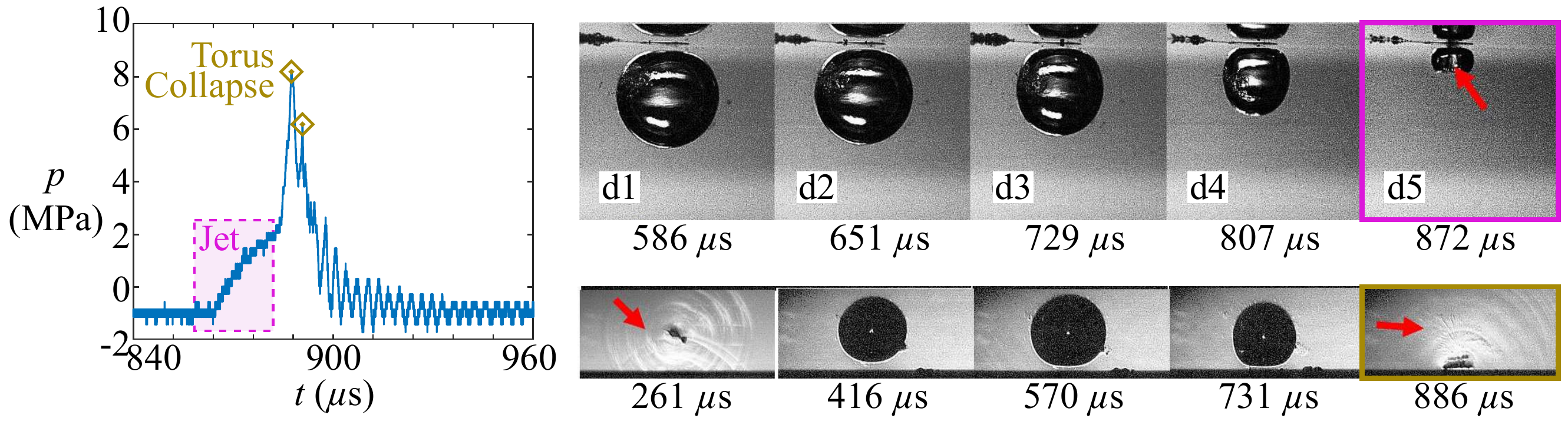}} \\
  \subfloat[maximum pressure at the wall during the first collapse]{\vspace{-0.2cm}\includegraphics[height=0.325\textwidth]{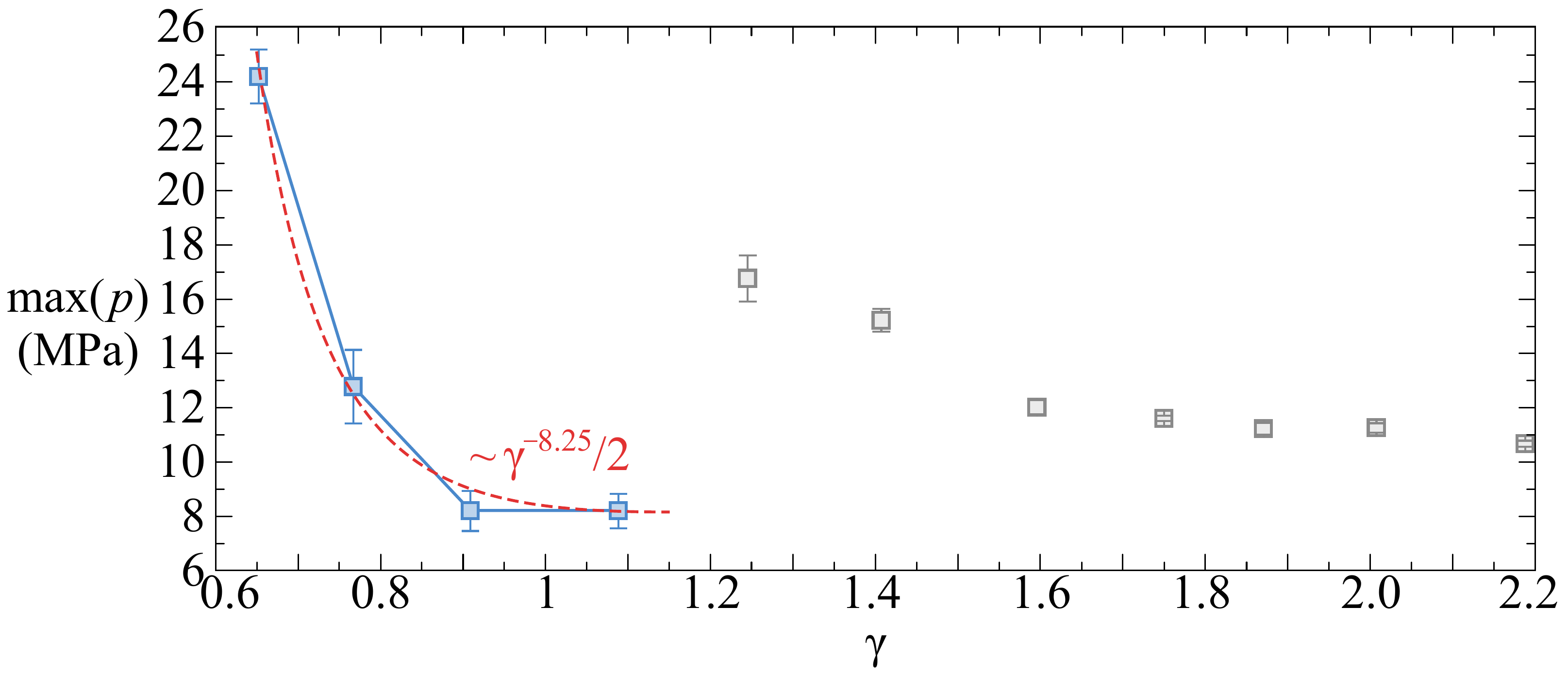}}
\caption{First regime for $\gamma \in [0.65,\ 1.09]$: (a--d) Time evolution of wall pressure (left panels), microjet visualizations (top right row), shock wave visualizations (bottom right row). (e) Maximum pressure at the wall (markers) corresponding with power-law fit (dashed).}
\label{fig:TorusCollapse}
\end{figure}

Figure \ref{fig:TorusCollapse}(a--d) depicts the wall pressure (left panel), the bubble interface (top 5-figures row) and the shock wave visualization (bottom 5-figures row) for $\gamma \leq 1.09$. For such low stand-off distances, the microjet occurs before the bubble collapse. This is a key observation as the microjet is necessary to the emergence of the torus collapse as the jet pierces the bubble before the actual collapse occurs. Hence, the jet itself does not produce the largest pressure spike, however it controls the bubble topology, leading to the resulting toroidal shock-wave source. According to \cite{ShockPressureDistance}, the pressure of the shock wave is inversely proportional to the distance between the shock source and the rigid wall. Moreover, when the shock waves are moving in the fluid, a residual internal energy will be left behind based on the increment in the entropy behind the shock front \citep{Brinkley1947}. This leads to the reduction in the observed shock energy when the wall sensor gets farther from the shock source. 

By observing the shock wave and considering the wall pressure signal depicted in fig.\ \ref{fig:TorusCollapse}(a--d), the toroidal bubble collapse emits a quasi-spherical shock wave that interacts with itself on the wall when the two shock fronts meet. They then tend to amplify the resulting pressure at the wall as observed in the pressure sensor measurements (see multiple peaks in the left panel of fig.\ \ref{fig:TorusCollapse}(a--d)). Moving from larger to smaller $\gamma$, the centroid of the toroidal bubble approaches the wall. Additionally, the shock wave velocity is higher near the source and decreases exponentially to the speed of sound in water as $\gamma$ decreases \citep{ShockVelocity}. This causes the first self-interaction point of the shock waves to move closer to the wall. Our camera visualizations support this finding (not shown). Such observation is consisitent with the consideration that for $\gamma = 0.65$, multiple pressure peaks at larger $\gamma$ merge into a single intense peak at the wall, which our sensor cannot fully resolve due to its quick dynamics. As $\gamma$ decreases, the microjet occurs earlier in the collapse phase, affecting the inertia of the bubble interface less and resulting in more energy radiated as shock waves. Consequently, the maximum pressure increases as $\gamma$ decreases.

Figure \ref{fig:TorusCollapse}(e) depicts in light-blue markers the maximum wall pressure observed for $\gamma\leq 1.09$. As the pressure emitted by the shock is supposed to scale as a power-law in $\gamma$, the maximum pressure produced by the torus collapse is supposed itself to follow a power-law in $\gamma$, decreasing upon an increase of $\gamma$. This is demonstrated by the red dashed line fitted to the experimental data for $\gamma\leq 1.09$. The good agreement between our power-law fit and the experimental data further confirms our previous arguments and the literature observations.

\begin{figure}
\centering
  \subfloat[$\gamma = 1.24$]{\vspace{-0.2cm}\includegraphics[height=0.25\textwidth]{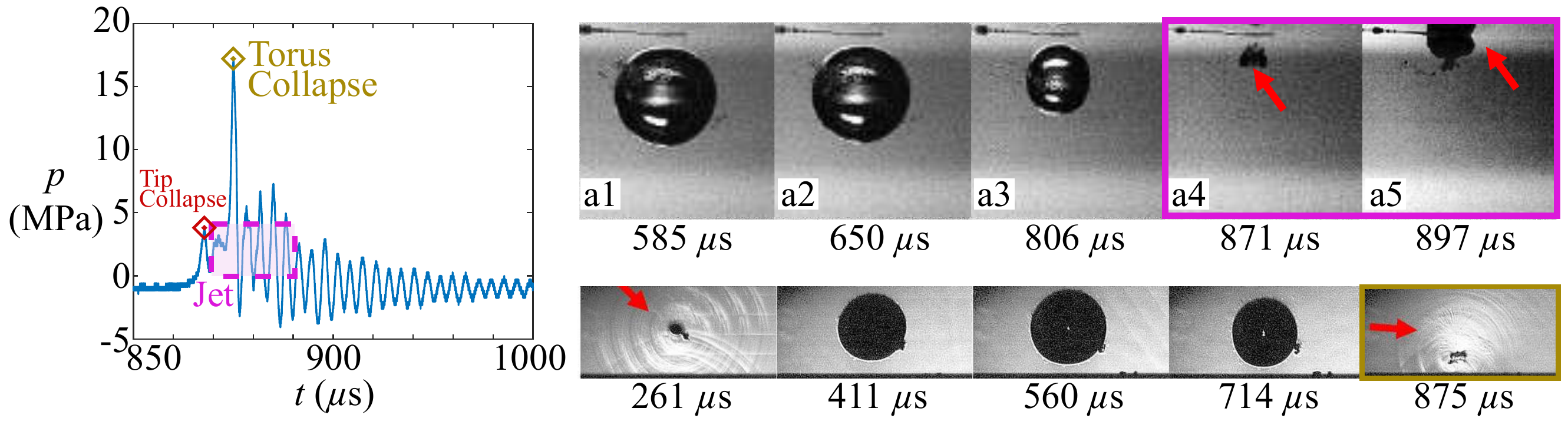}}\\
  \subfloat[$\gamma = 1.41$]{\vspace{-0.2cm}\includegraphics[height=0.25\textwidth]{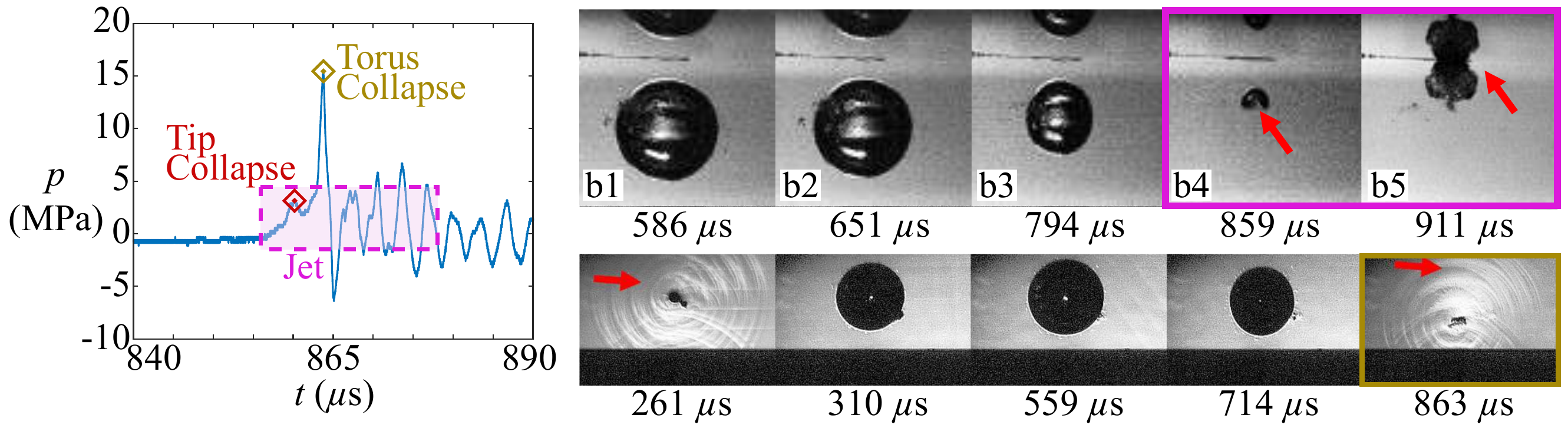}}\\
  \subfloat[$\gamma = 1.60$]{\vspace{-0.2cm}\includegraphics[height=0.25\textwidth]{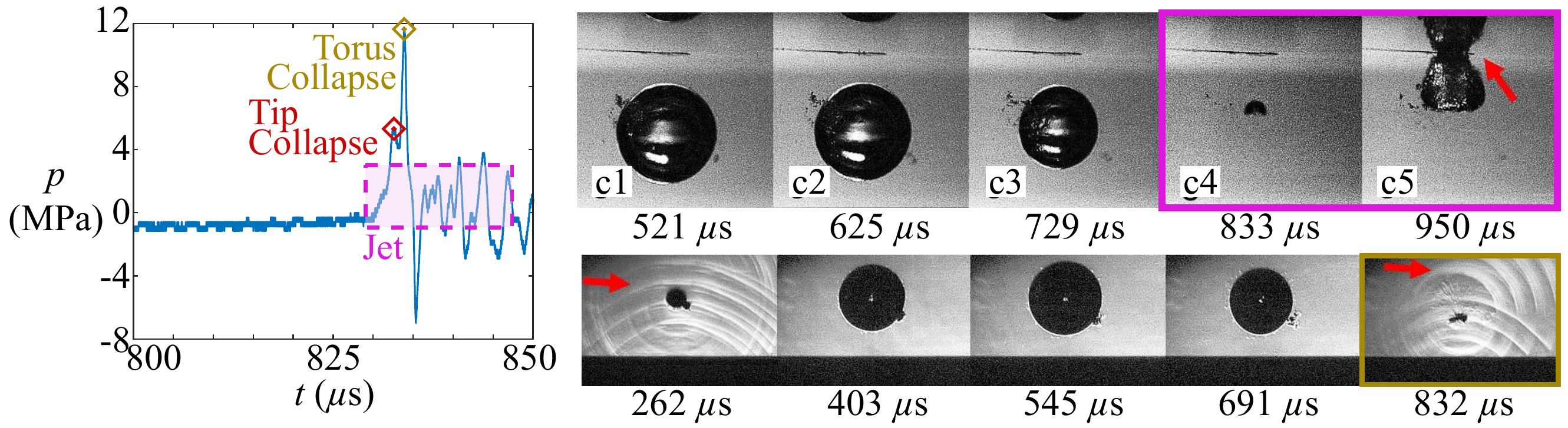}} \\
  \subfloat[maximum pressure at the wall during the first collapse]{\vspace{-0.2cm}\includegraphics[height=0.325\textwidth]{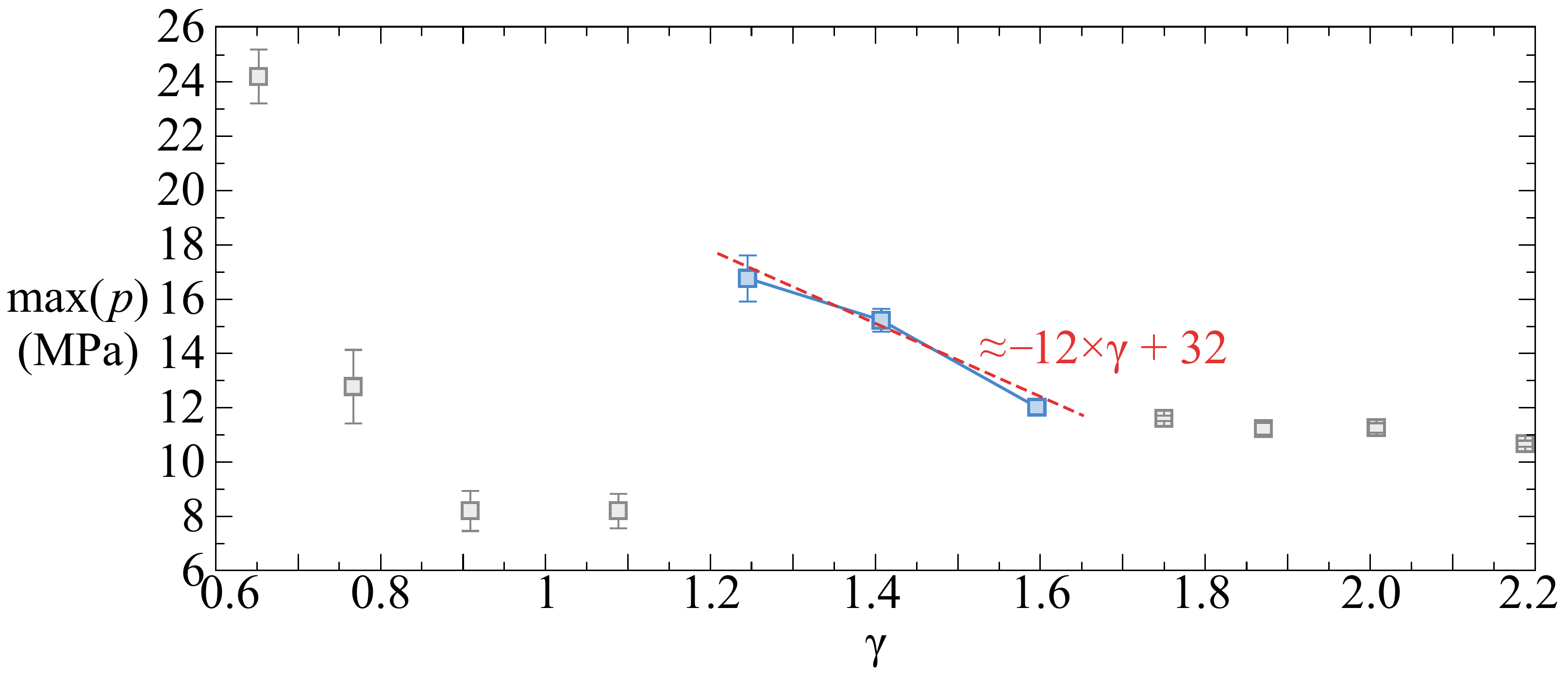}}
\caption{Second regime for $\gamma \in [1.24,\ 1.60]$: (a--c) Time evolution of wall pressure (left panels), microjet visualizations (top right row), shock wave visualizations (bottom right row). (d) Maximum pressure at the wall (markers) corresponding with linear fit (dashed).}
\label{fig:MixedCollapse}
\end{figure}

Upon an increase in stand-off ratio, hence passing from fig.\ \ref{fig:TorusCollapse} to fig.\ \ref{fig:MixedCollapse}, the first collapse phase starts with a tip collapse that occurs before the microjet pierces the bubble interface. Therefore, the microjet induces a topological change of the bubble interface by piercing the bubble \textit{after} the tip collapse. This prepares the interface topology for another collapse -- this time of toroidal kind (see ocher markers in the left panels of fig.\ \ref{fig:MixedCollapse}(a--c)) -- in immediate succession to the tip collapse (see red markers in the left panels of fig.\ \ref{fig:MixedCollapse}(a--c)). The torus collapse builds up pressure on the tip collapse, leading therefore to a significant increase of $\max(p)$ at the wall when passing from $\gamma = 1.09$ (purely torus collapse regime) to $\gamma = 1.24$ (mixed tip-and-torus collapse regime). This mixed regime is observed as a robust feature up to $\gamma = 1.60$. The tip collapse, however, occurs further away from the wall upon an increase of the stand-off ratio. This implies that the torus collapse has strongly interacted with the tip collapse dissipating part of its energy in shock refractions long before reaching the wall. As a result, the maximum pressure at the wall is decreased passing from $\gamma = 1.24$ to $\gamma=1.60$ (see fig.\ \ref{fig:MixedCollapse}(d)). In accordance with the experimental data (light-blue markers in fig.\ \ref{fig:MixedCollapse}(d)), we fitted linearly $\max(p)$ and $\gamma$ (red dashed line in fig.\ \ref{fig:MixedCollapse}(d)). Such a linear trend is not to be intended as derived from a theoretical prediction, but rather as a guide for the eyes that could help identifying the order of magnitude of $\max(p)$ once $\gamma$ is known based on the experimental evidence. 

\begin{figure}
\centering
  \subfloat[Bubble at $\max_{t} (R_{eq})$]{\includegraphics[height=0.30\textwidth]{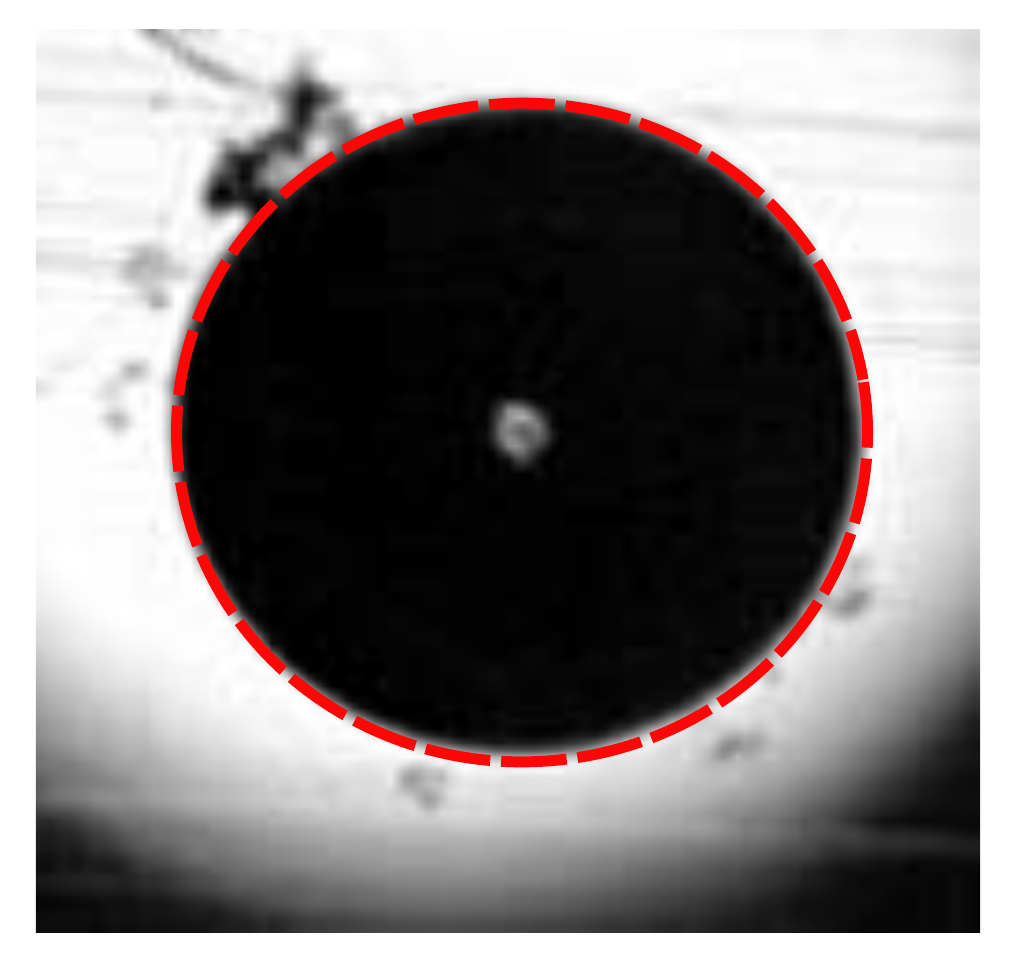}} \quad
  \subfloat[Bubble at $V_{tb}$]{\includegraphics[height=0.30\textwidth]{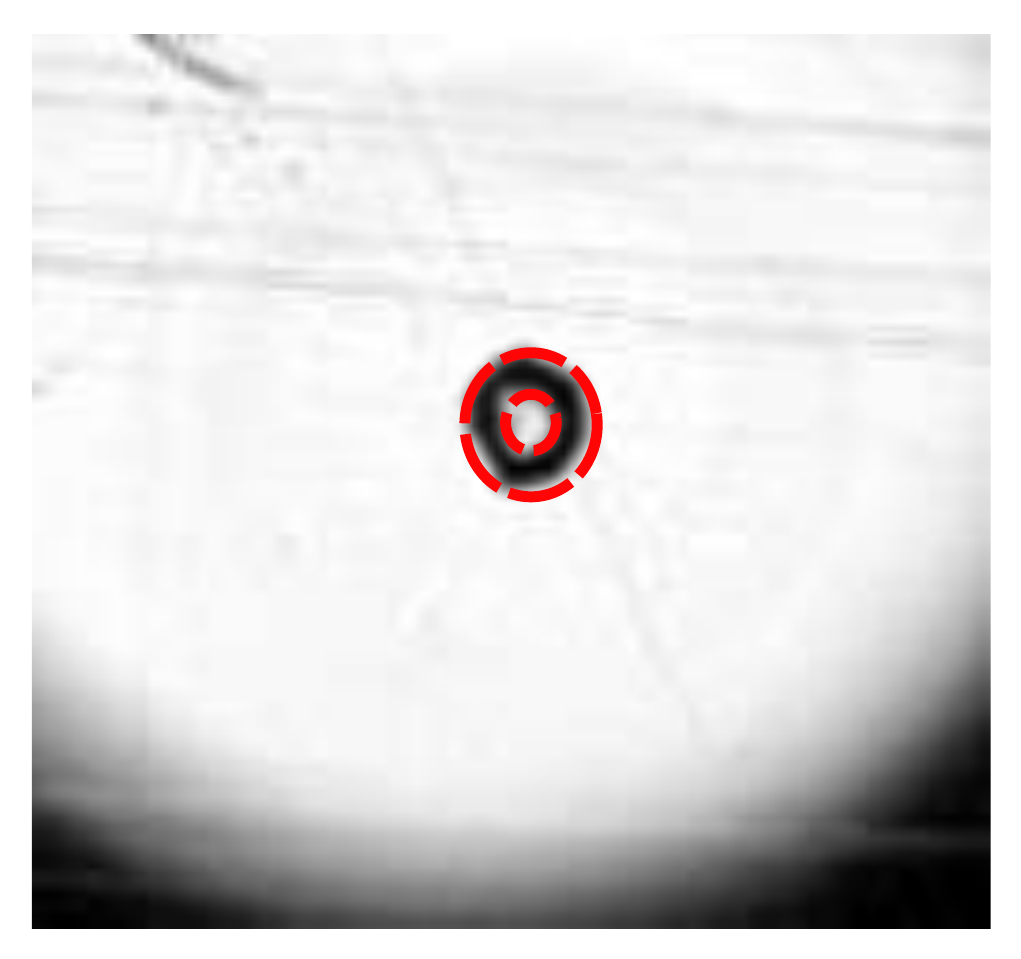}} \quad
\caption{(a) Cavitation bubble at its maximum radius for $\gamma$=1.64. (b) Bubble at its minimum toroidal volume at which the shock wave gets emitted during compression for $\gamma$=1.64.}
\label{fig:Compression}
\end{figure}

The second regime identified for $\gamma \in [1.24,\ 1.60]$ and depicted in fig.\ \ref{fig:MixedCollapse} is the most intricate dynamics among the cases we consider. As for the first regime (\textit{purely torus collapse}, see fig.\ \ref{fig:TorusCollapse}), also for the \textit{mixed tip-and-torus regime}, the torus collapse corresponds to the highest pressure peak, as reported in fig.\ \ref{fig:TorusCollapse}(e) and fig.\ \ref{fig:MixedCollapse}(d). An estimate of the shock-wave intensity resulting from a torus collapse is provided by the rate of compression $\Sigma$ proposed by \cite{Lindau2003}. This indicator considers the ratio between the volume of the bubble when it reaches its maximum radius $V_\text{max} = V(t = t_{MR})$ and the volume of the smallest toroidal bubble $V_\text{tb}$ right before torus collapse, i.e. $\Sigma = V_\text{max}/V_\text{tb}$. We stress that the \textit{smallest} toroidal bubble is used to calculate $V_\text{tb}$, which does \textit{not} correspond to the volume of the toroidal bubble formed right after jet pierced the bubble interface. In fact, the bubble will continue to shrink for sometime until the torus collapse occurs. In our study, $V_\text{tb}$ is estimated by the frame emitting the torus shock wave as the smallest toroidal bubble. In order to estimate $\Sigma$ by using our experimental visualizations, the smallest toroidal bubble volume is calculated from the front view of the bubble interface (see fig.\ref{fig:Compression}). Although the pierced bubble is not a perfect toroid, it is assumed to be a toroidal bubble which is a practice accepted in the literature to estimate $V_\text{tb}$ \vspace{1 cm}. 

\begin{figure}
  \centerline{\includegraphics[height=0.5\textwidth]{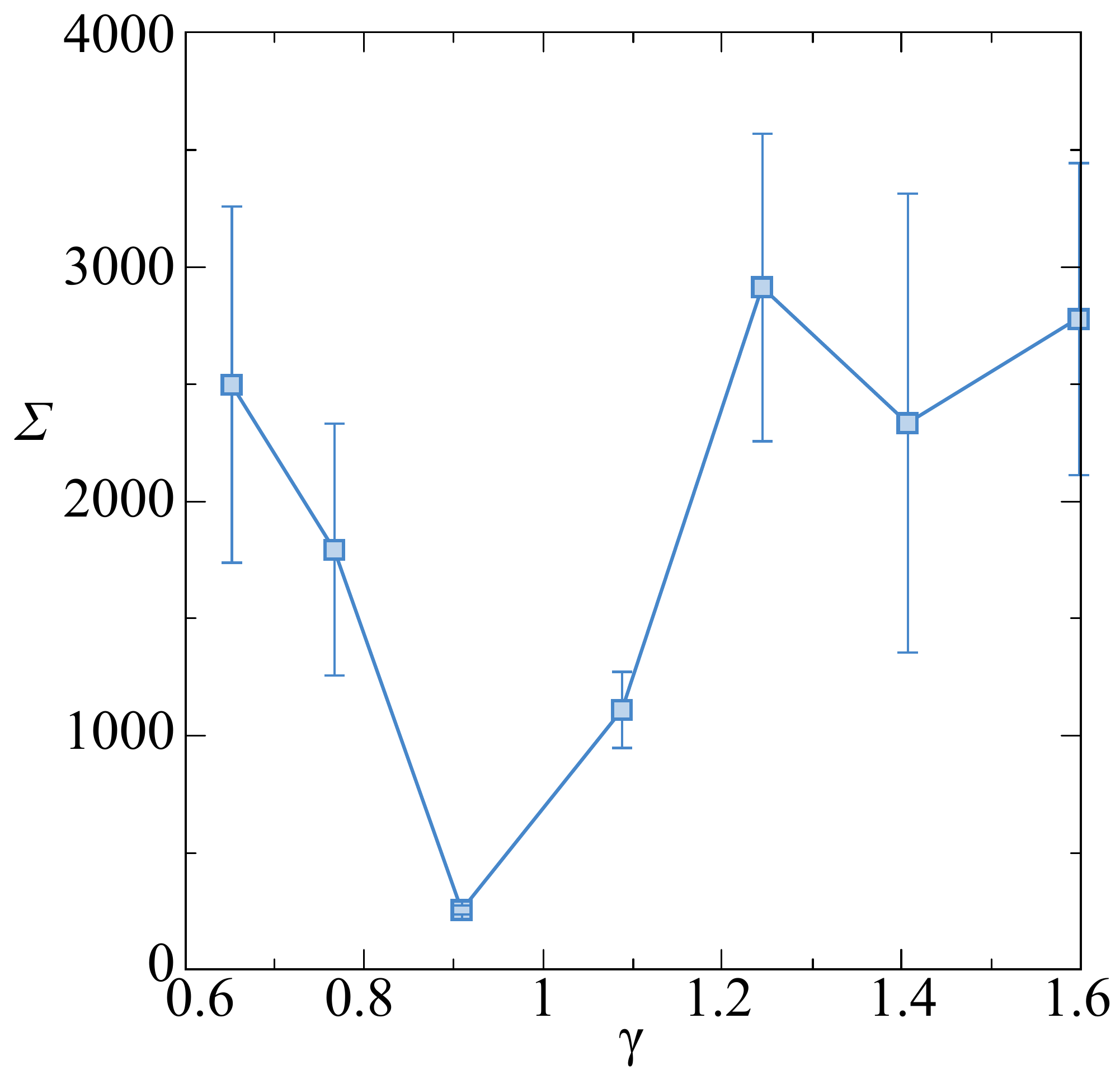}}
  \caption{Experimental estimation of the compression ratio $\Sigma$ for the torus bubble collapse.}
\label{fig:CompressioRatio}
\end{figure}

A corresponding plot of the compression ratio $\Sigma$ is depicted in fig.\ \ref{fig:CompressioRatio}. High values of $\Sigma$ signify that the bubble built up a significant energy between its point of maximum volume $V_\text{max}$ and the corresponding point of torus collapse at $V_\text{tb}$. Such an energy will be released in the torus shock-wave, hence $\Sigma$ is expected to control the $\max(p)$ of torus collapses. This is demonstrated by comparing fig.\ \ref{fig:CompressioRatio} with fig.\ \ref{fig:TorusCollapse}(e) and fig.\ \ref{fig:MixedCollapse}(d), which allows us to conclude that the transition between the torus and mixed collapse regimes identified for $\gamma\leq 1.09$ and $\gamma\in [1.24,\ 1.60]$, respectively, can be understood by explaining the trend of $\Sigma$. We therefore speculate that the presence of a tip collapse right prior the torus collapse helped reducing the minimum volume before the onset of the torus collapse for $\gamma\in [1.24,\ 1.60]$. In turn, reducing $V_\text{tb}$ this led to a higher $\Sigma$, hence to an increase of $\max(p)$ passing from $\gamma = 1.09$ to $\gamma = 1.24$.

\begin{figure}
\centering
  \subfloat[$\gamma = 1.75$]{\vspace{-0.2cm}\includegraphics[height=0.25\textwidth]{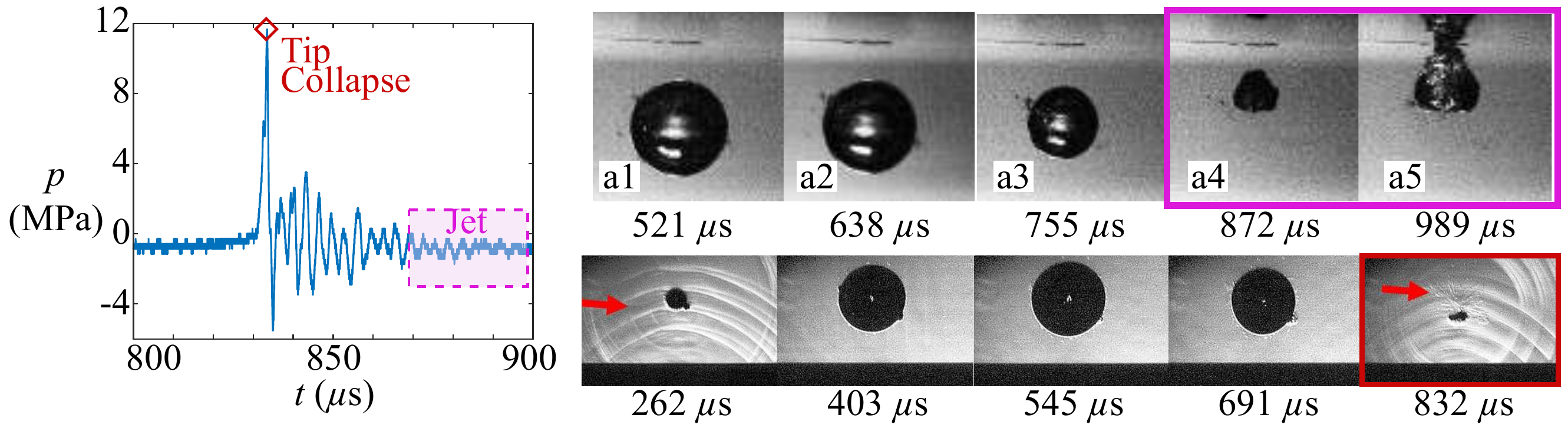}} \\
  \subfloat[$\gamma = 1.87$]{\vspace{-0.2cm}\includegraphics[height=0.25\textwidth]{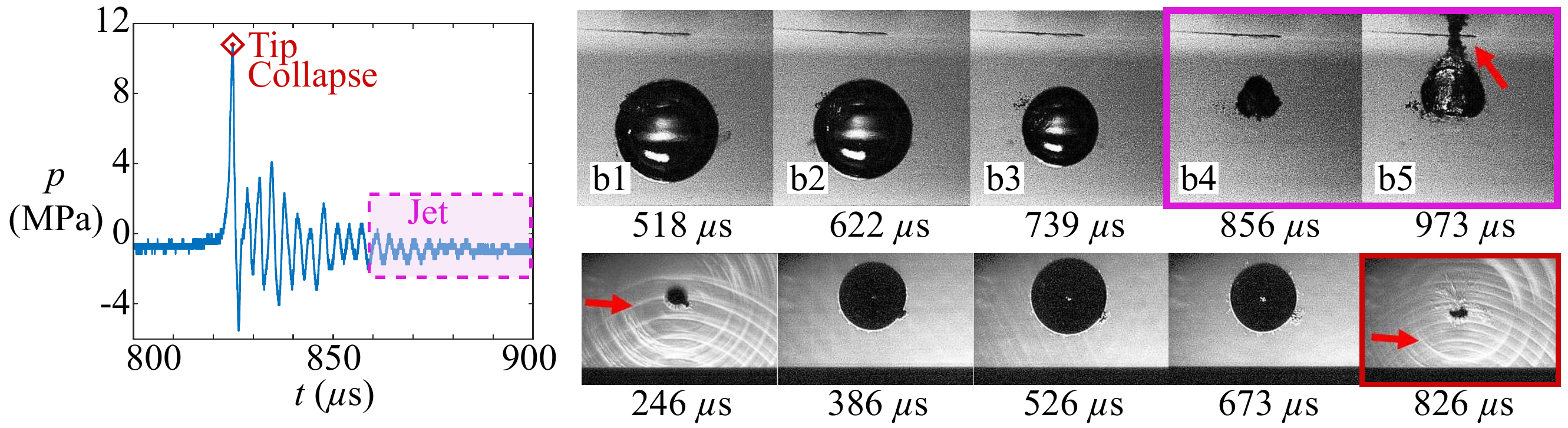}} \\
  \subfloat[$\gamma = 2.01$]{\vspace{-0.2cm}\includegraphics[height=0.25\textwidth]{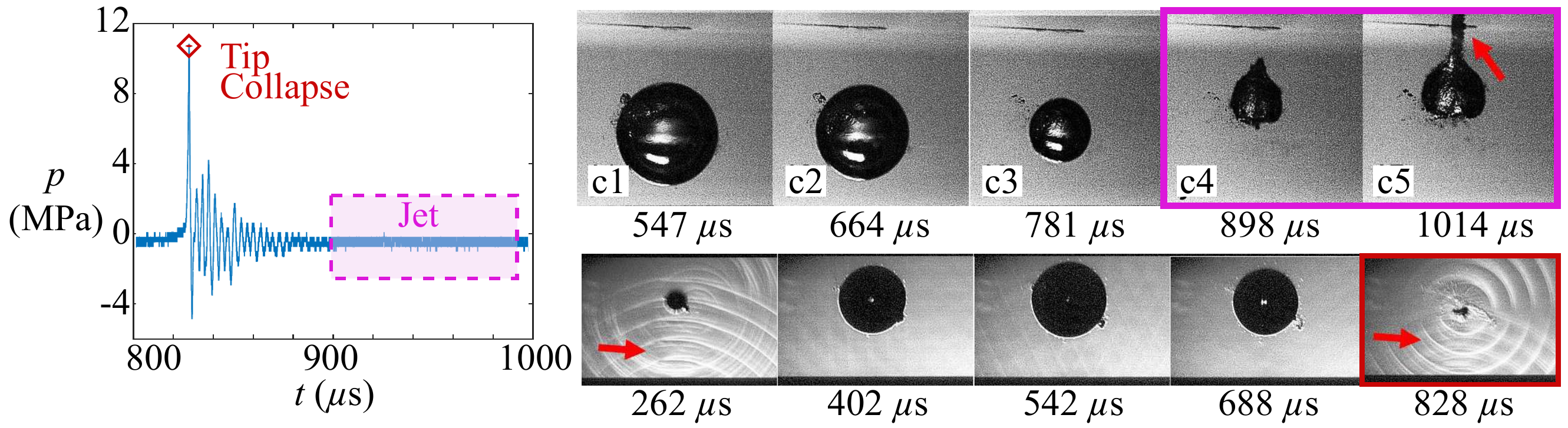}} \\
  \subfloat[$\gamma = 2.19$]{\vspace{-0.2cm}\includegraphics[height=0.25\textwidth]{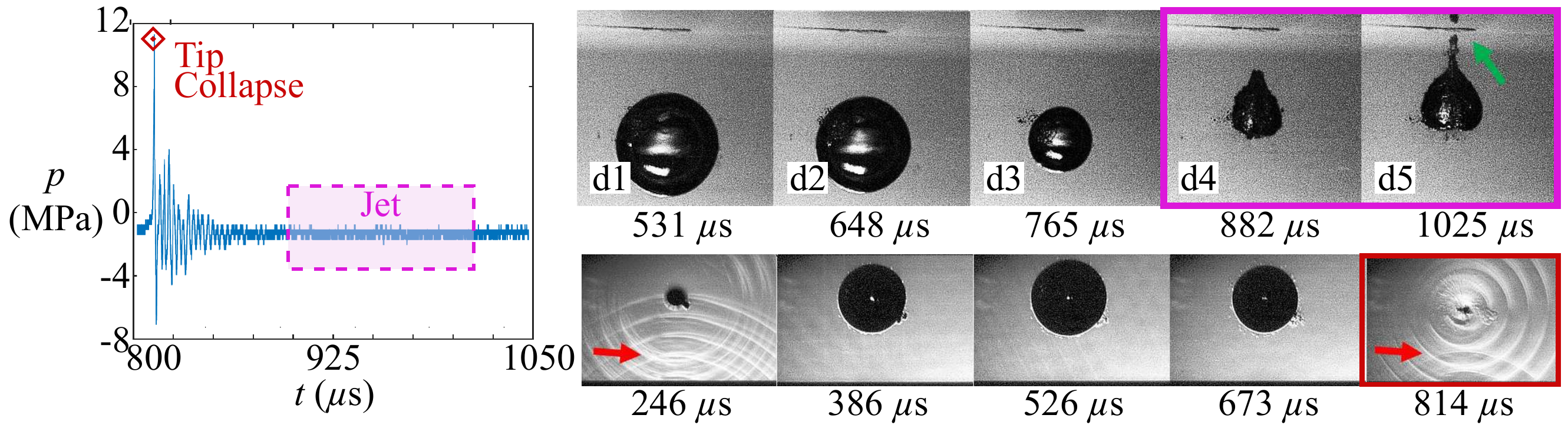}} \\
  \subfloat[maximum pressure at the wall during the first collapse]{\vspace{-0.2cm}\includegraphics[height=0.325\textwidth]{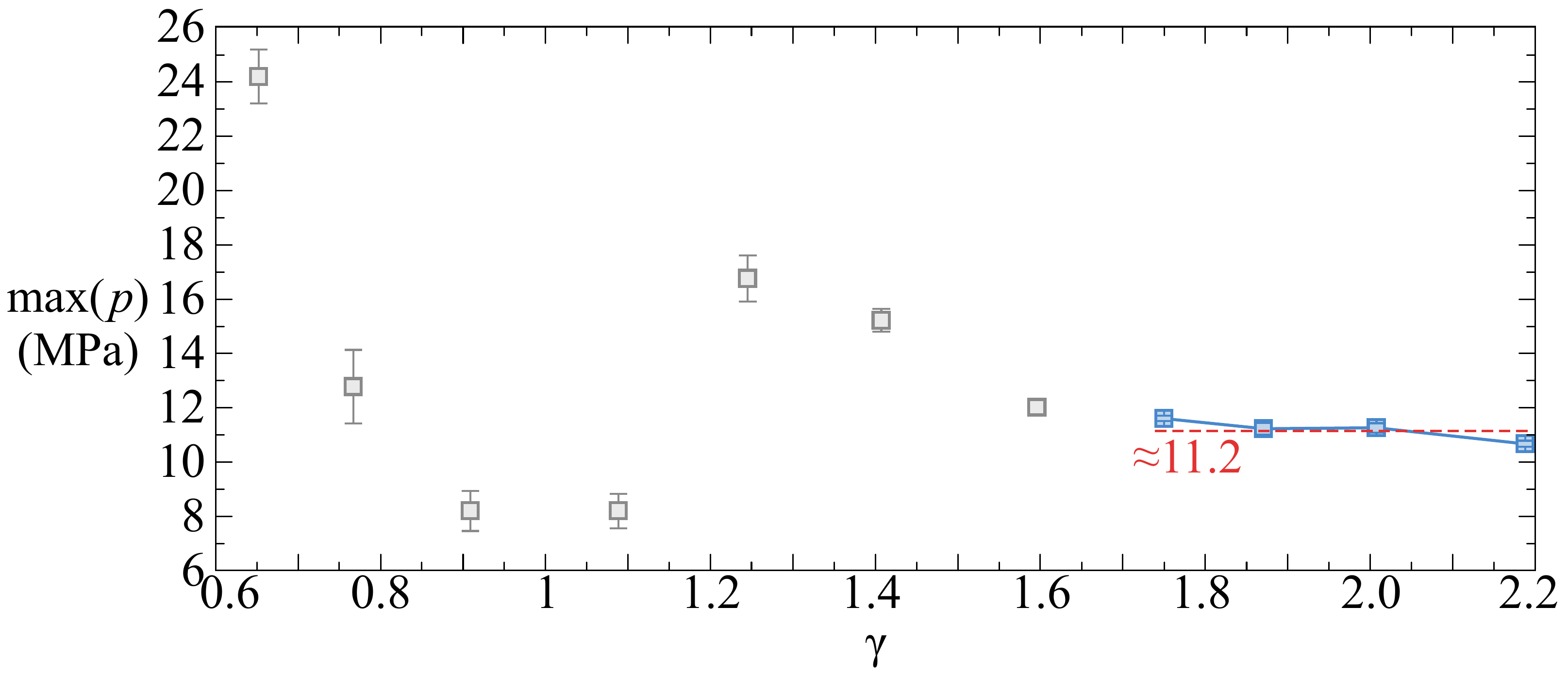}}
\caption{Third regime for $\gamma \in [1.75,\ 2.19]$: (a--d) Time evolution of wall pressure (left panels), microjet visualizations (top right row), shock wave visualizations (bottom right row). (e) Maximum pressure at the wall (markers) corresponding with constant fit (dashed).}
\label{fig:TipCollapse}
\end{figure}

Finally, a third regime for the first collapse is observed upon an increase of the stand-off ratio (see fig.\ \ref{fig:TipCollapse}). As for $\gamma \in [1.24,\ 1.60]$, also when the stand-off ratio ranges between $\gamma \in [1.75,\ 2.19]$ the tip collapse occurs before the microjet pierces the bubble. The bubble is now far enough from the wall to significantly delay the occurrence of the microjet. As a result, the bubble does not get pierced by the microjet immediately after tip collapse, hence it does not undergo a quick successive collapse (toroidal in kind as for the mixed regime, see left panels of fig.\ \ref{fig:TipCollapse}(a--d)). The maximum pressure at the wall closely resemble the one corresponding to a spherical shock wave, which is not much affected by $\gamma$ (see fig.\ \ref{fig:TipCollapse}(c)). This specific point will be further proven by numerical simulations in our follow-up papers (Part 2 and 3 of this study). A \textit{purely tip collapse} regime is therefore identified for $\gamma \geq 1.75$, as the microjet becomes therefore negligible for describing the wall pressure during the first collapse phase.

\subsection{Wall pressure during the second bubble collapse}
\vspace{5pt}

After enduring the initial collapse, the bubble retains some of its energy without fully radiating it away. This results in subsequent rebounds and followed by successive collapses of the bubble. While there can be multiple rebounds following the inertial collapse of the bubble, we focus here only on the first rebound, which is followed by the second collapse of the bubble. During this second collapse, we can still observe toroidal bubble formations, as discussed in the previous section. However, the second collapses further away from the wall exhibit different dynamics, that are characterized by \textit{hour-glass bubbles} and \textit{near-wall detached bubbles}, with no toroidal formations evident. 

\begin{figure}
\centering
  \subfloat[Toroid]{\includegraphics[height=0.29\textwidth]{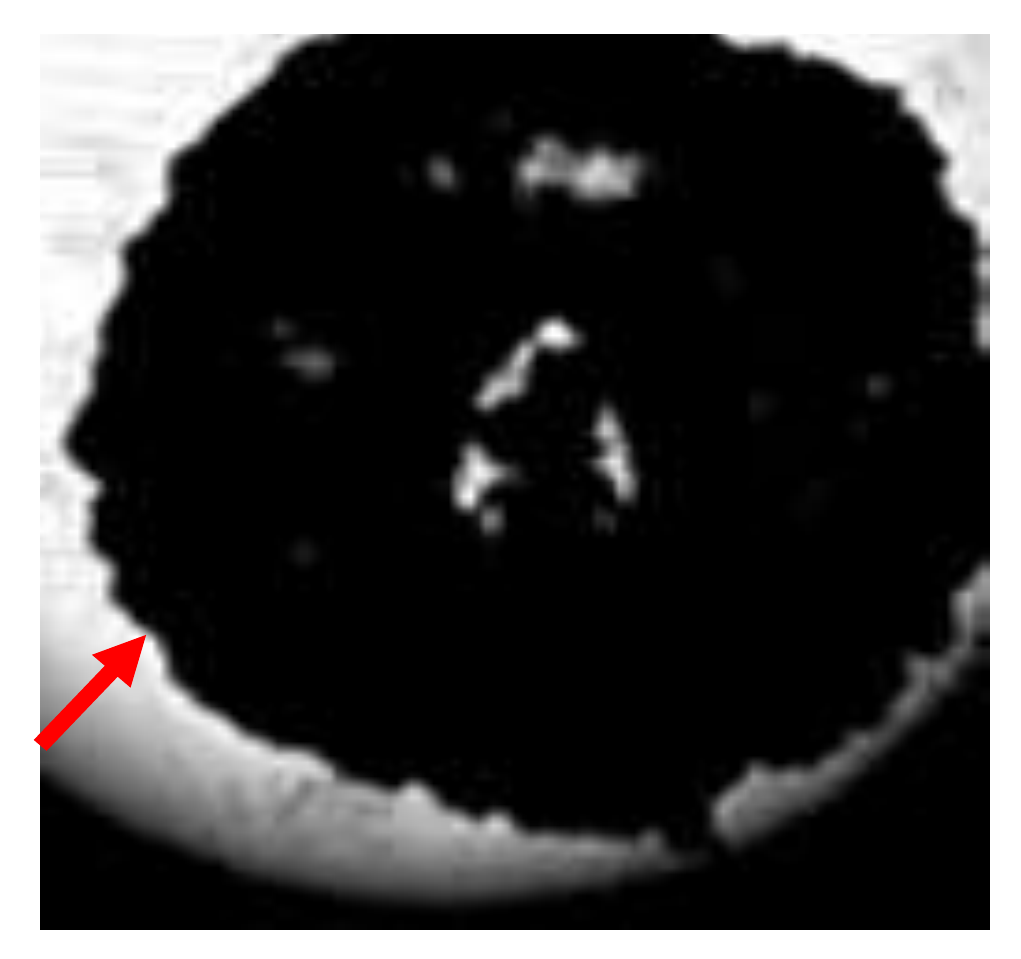}} \quad
  \subfloat[STF collapse]{\includegraphics[height=0.29\textwidth]{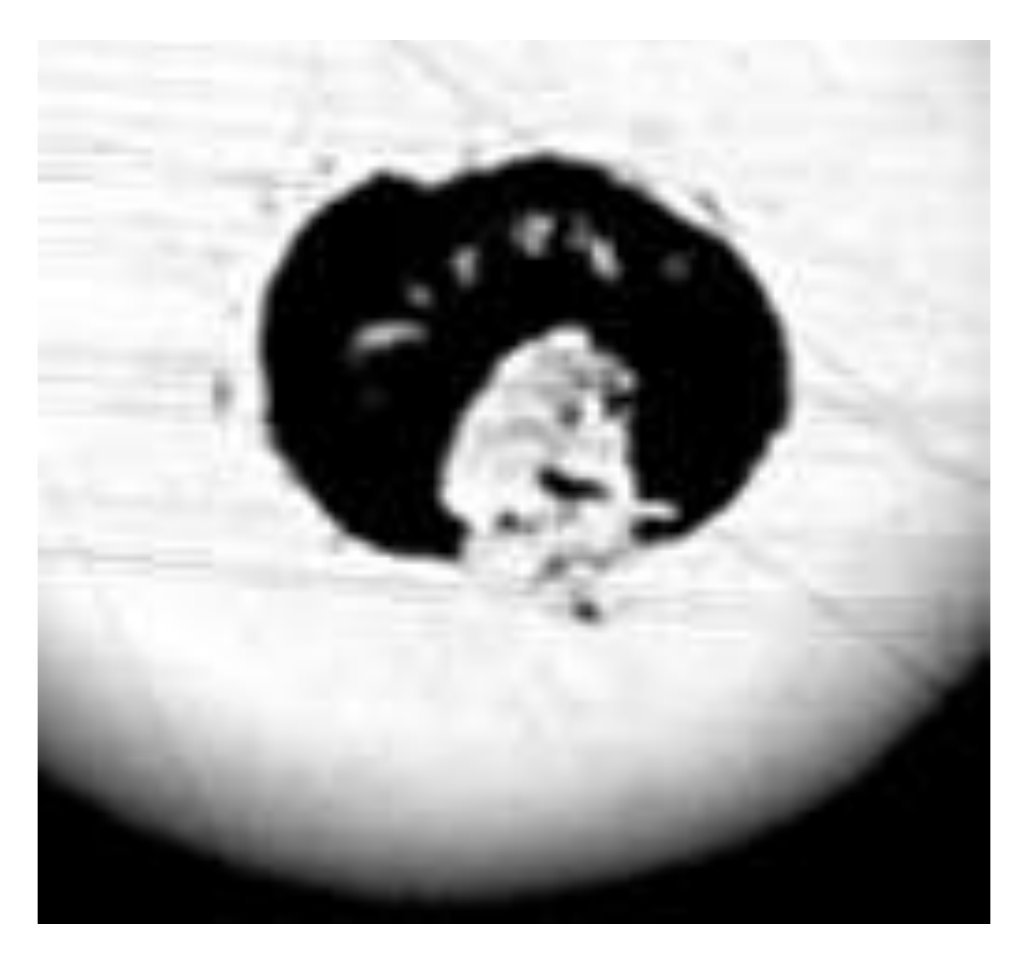}} \quad
  \subfloat[PTF collapse]{\includegraphics[height=0.31\textwidth]{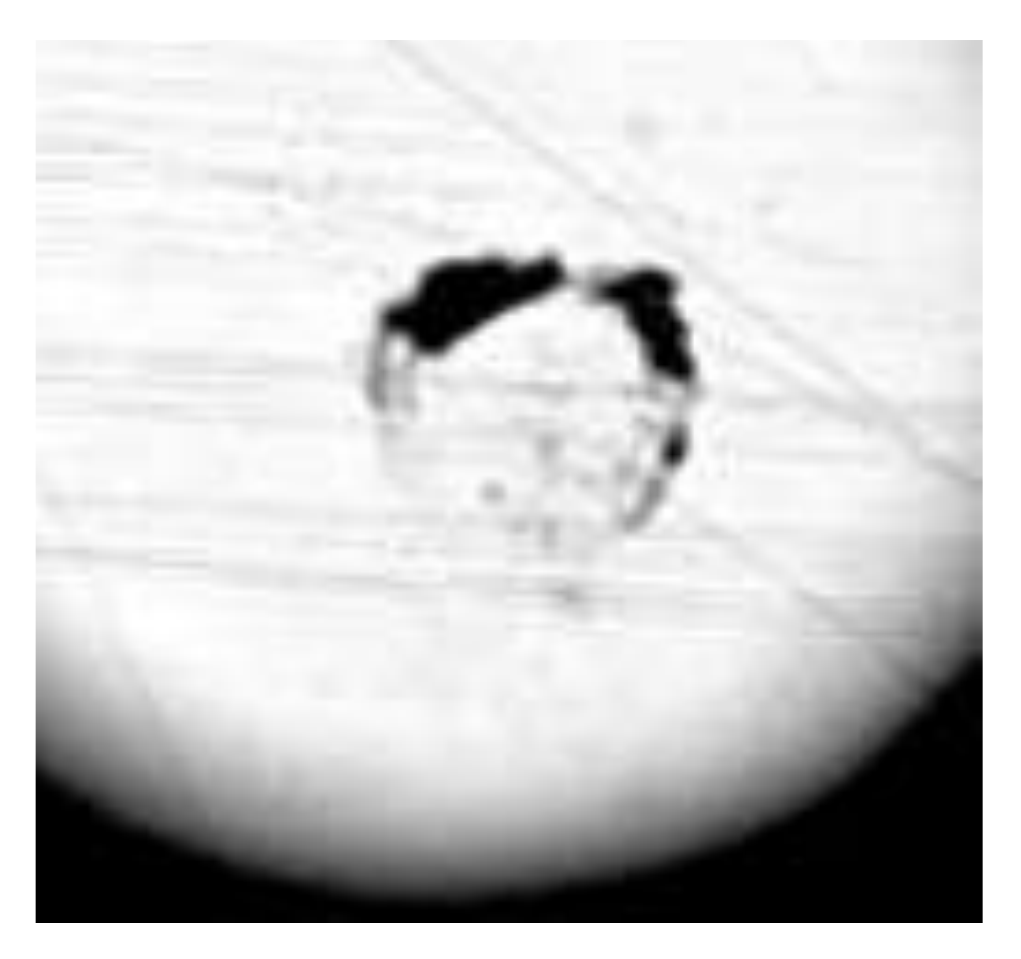}}
\caption{(a) Rebounded bigger toroidal bubble for $\gamma$=1.09. Possible Rayleigh-Taylor instability (red arrow) (b) STF collapse emitting shock wave, also showing the actual PTF. (c) PTF collapse emitting shock wave.}
\label{fig:PTFandSTF}
\end{figure}

The first two regimes ($\gamma \leq 1.09$, and $\gamma \in [1.24,\ 1.60]$) lead to the formation of toroidal bubbles also for the second collapse. However, these toroidal bubbles do not collapse in one piece. Instead, they dissociate into fragments that we name as the \textit{primary toroidal fragment} (PTF) and the \textit{secondary toroidal fragment} (STF). The STF represents the dissociation of the lower part of the toroidal bubble, while the PTF refers to the upper part of the toroidal bubble (see fig. \ref{fig:PTFandSTF}). 

\begin{figure}
\centering
  \subfloat[$\gamma = 0.65$]{\vspace{-0.2cm}\includegraphics[height=0.25\textwidth]{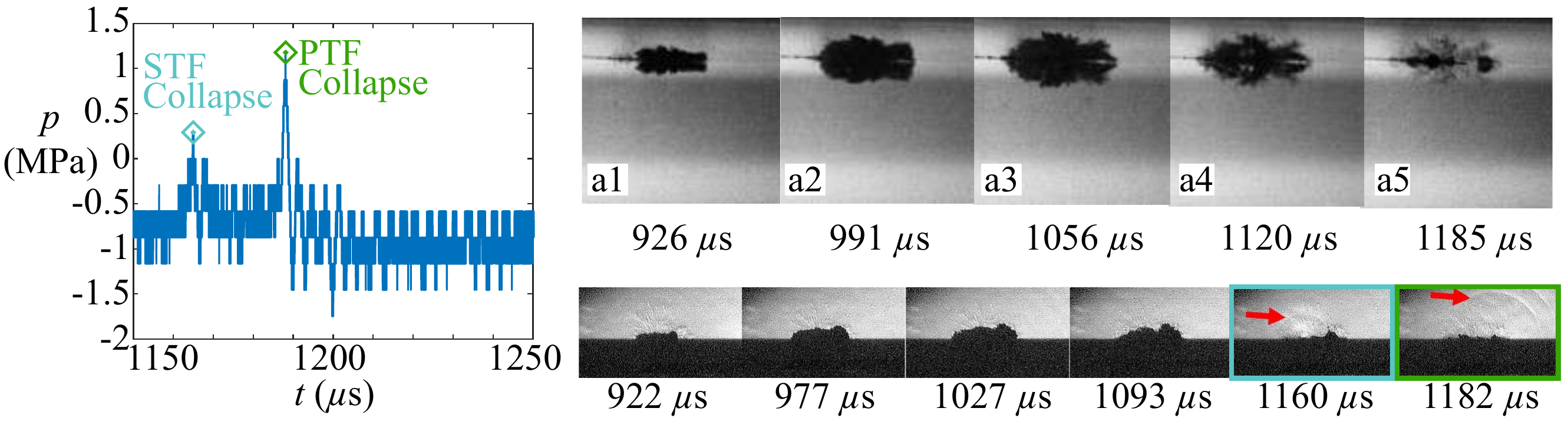}} \\
  \subfloat[$\gamma = 0.77$]{\vspace{-0.2cm}\includegraphics[height=0.25\textwidth]{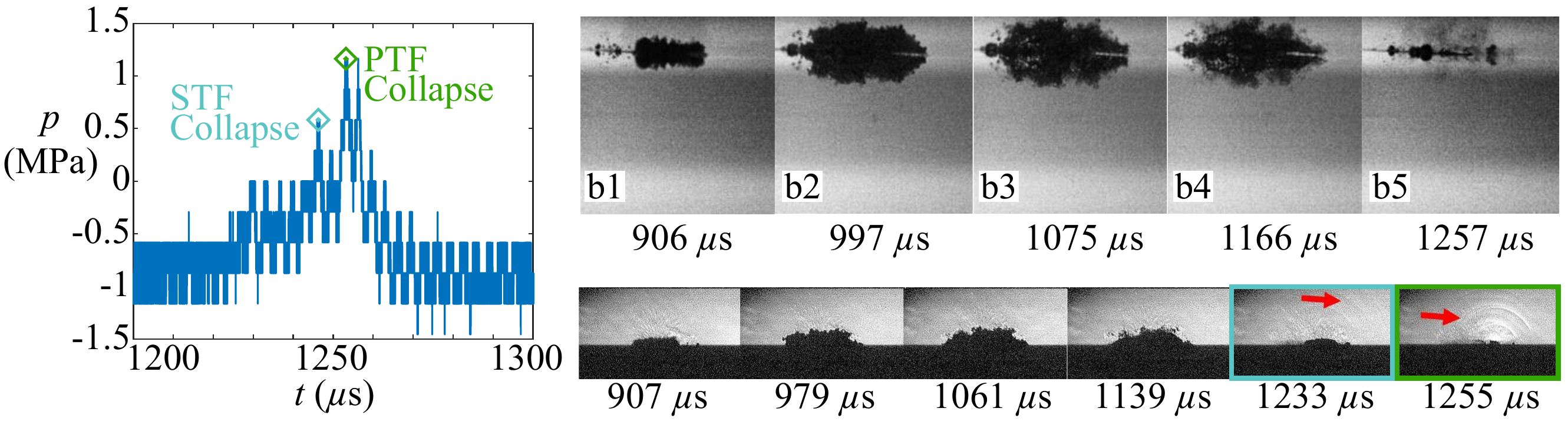}} \\
  \subfloat[$\gamma = 0.91$]{\vspace{-0.2cm}\includegraphics[height=0.25\textwidth]{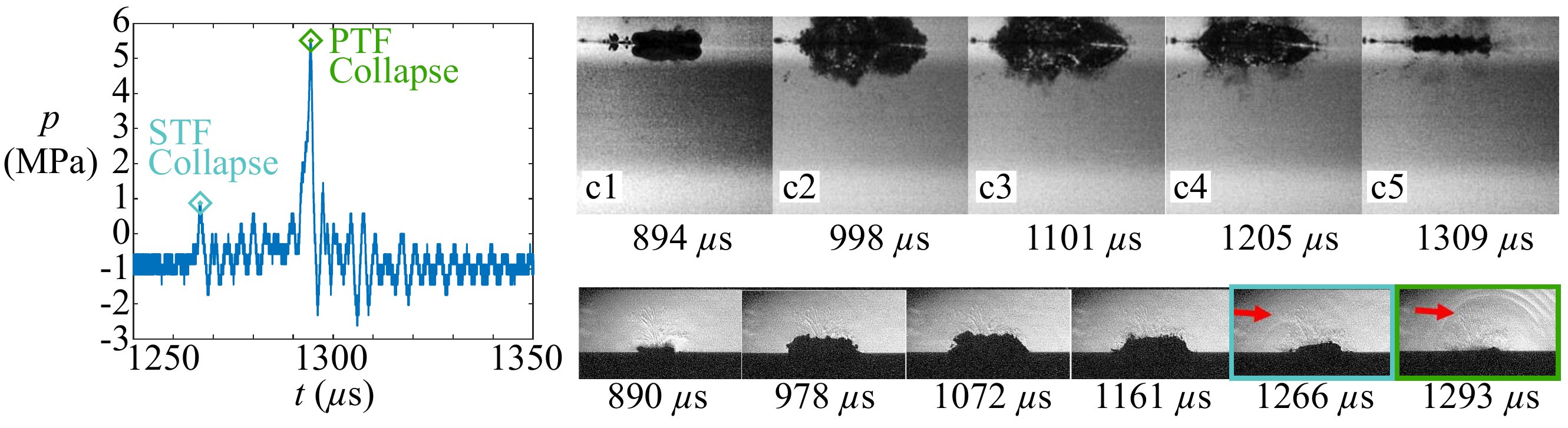}} \\
  \subfloat[$\gamma = 1.09$]{\vspace{-0.2cm}\includegraphics[height=0.25\textwidth]{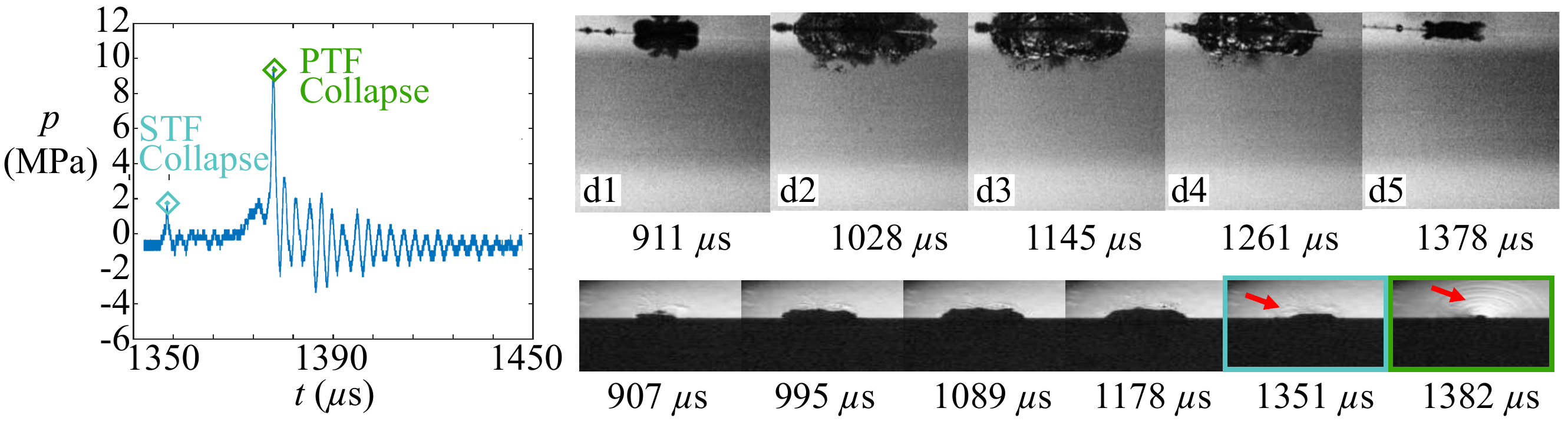}} \\
  \subfloat[maximum pressure at the wall during the second collapse]{\vspace{-0.2cm}\includegraphics[height=0.325\textwidth]{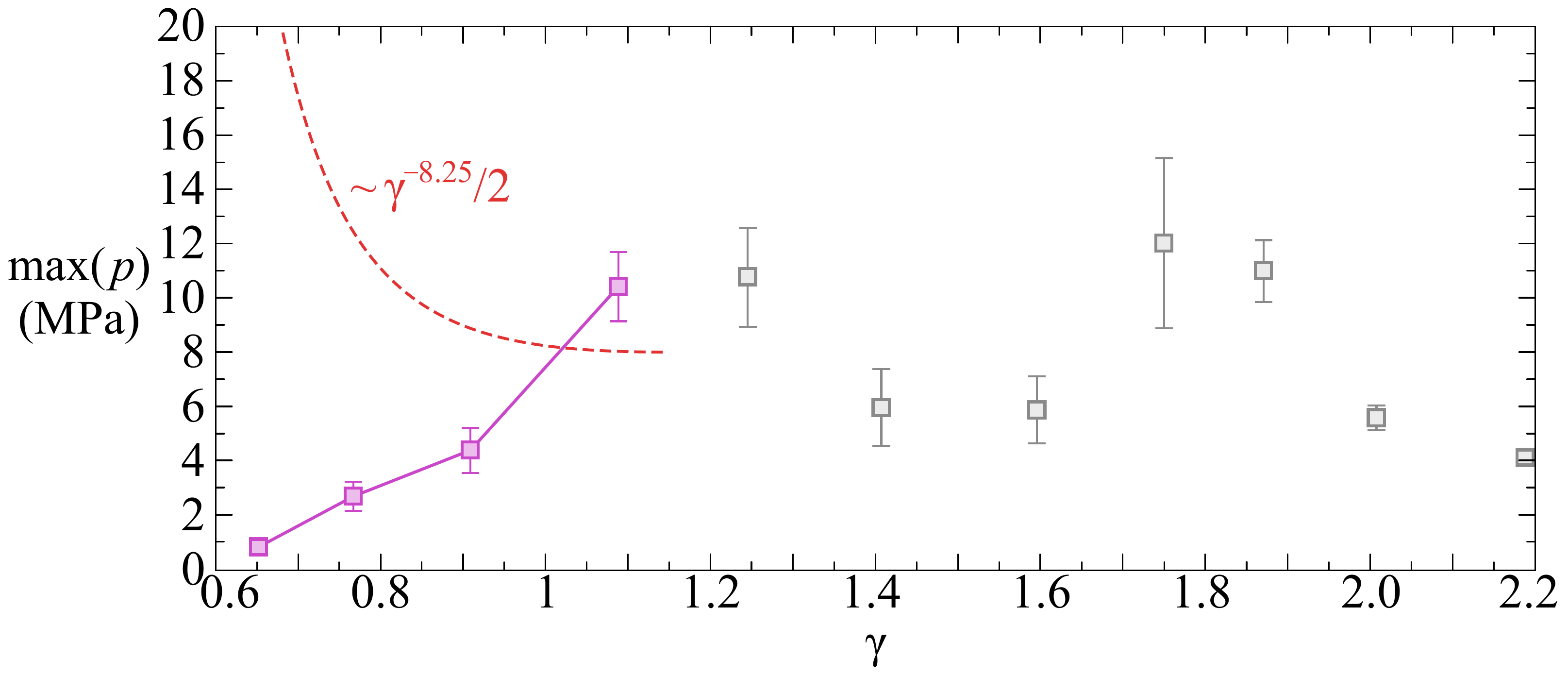}}
\caption{First regime for $\gamma \in [0.65,\ 1.09]$: (a--d) Time evolution of wall pressure (left panels), microjet visualizations (top right row), shock wave visualizations (bottom right row). (e) Maximum pressure at the wall (markers) corresponding with power-law fit (dashed).}
\label{fig:SecondCollapseTorusFirst}
\end{figure}

 Figure \ref{fig:SecondCollapseTorusFirst}(e) displays the pressure magnitudes from the second collapse (magenta markers) alongside the power-law fit observed during the first collapse. As discussed earlier, the timing of the jet occurrence during the final stages of the first collapse is crucial for the maximum wall pressure. In the first regime with larger $\gamma$ (from 1.09), the occurrence of the jet later during the first collapse allows the bubble to retain its maximum energy owing to the inertial disturbance. Consequently, the subsequent expansion of the rebounding bubble is more pronounced for larger $\gamma$ (from 1.09), and diminishes as $\gamma$ decreases (not shown). These rebounds result in the formation of a toroidal bubble, whose collapse mechanism first involves the collapse of the STF, followed by the collapse of the PTF (see light-blue, and green markers in left panels, and the corresponding shock images of fig. \ref{fig:SecondCollapseTorusFirst}(a-d)).

\begin{figure}
\centering
    \subfloat[$\gamma = 1.24$]{\vspace{-0.2cm}\includegraphics[height=0.25\textwidth]{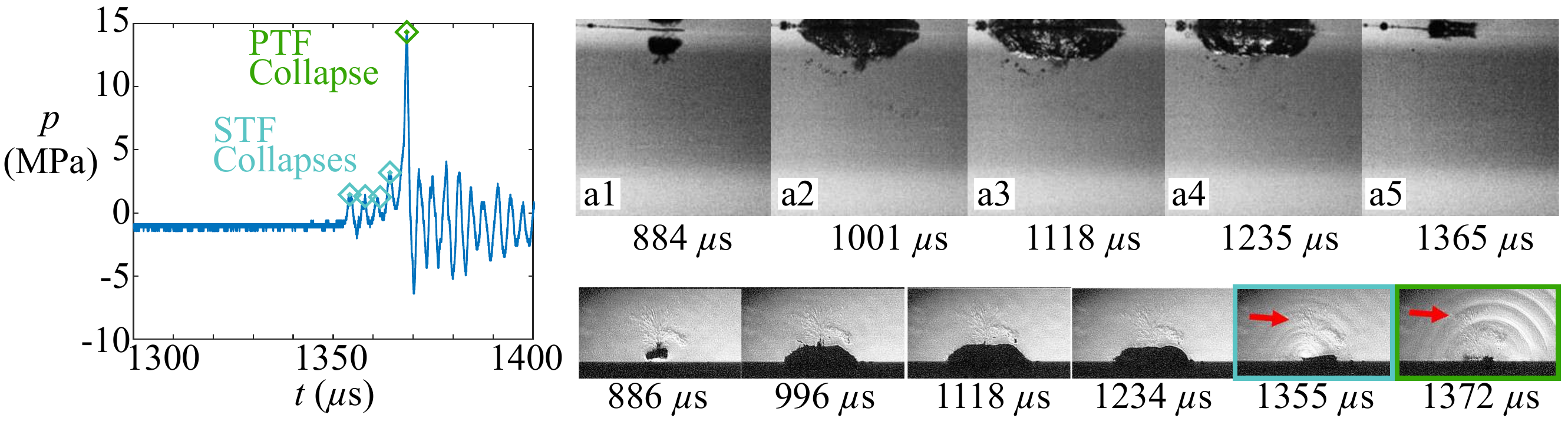}}\\
  \subfloat[$\gamma = 1.41$]{\vspace{-0.2cm}\includegraphics[height=0.25\textwidth]{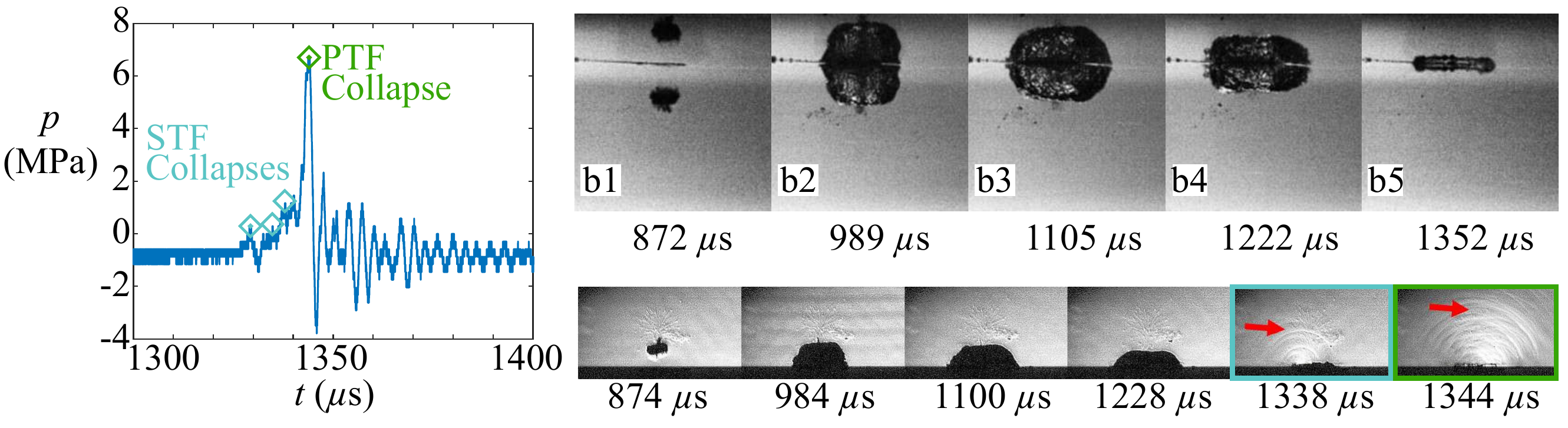}}\\
  \subfloat[$\gamma = 1.60$]{\vspace{-0.2cm}\includegraphics[height=0.25\textwidth]{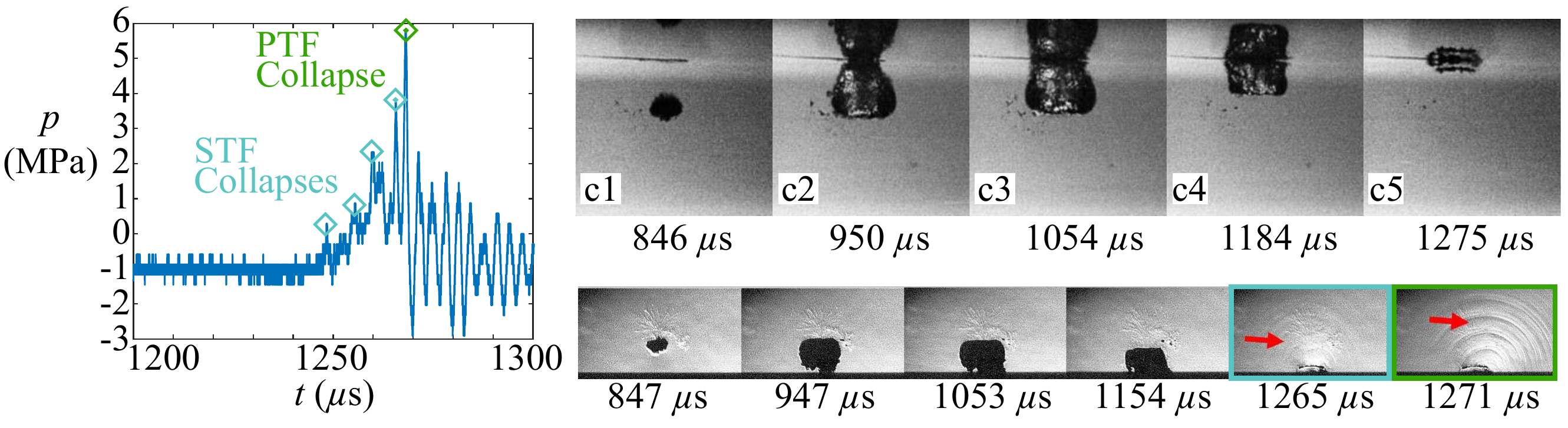}} \\
  \subfloat[maximum pressure at the wall during the second collapse]{\vspace{-0.2cm}\includegraphics[height=0.325\textwidth]{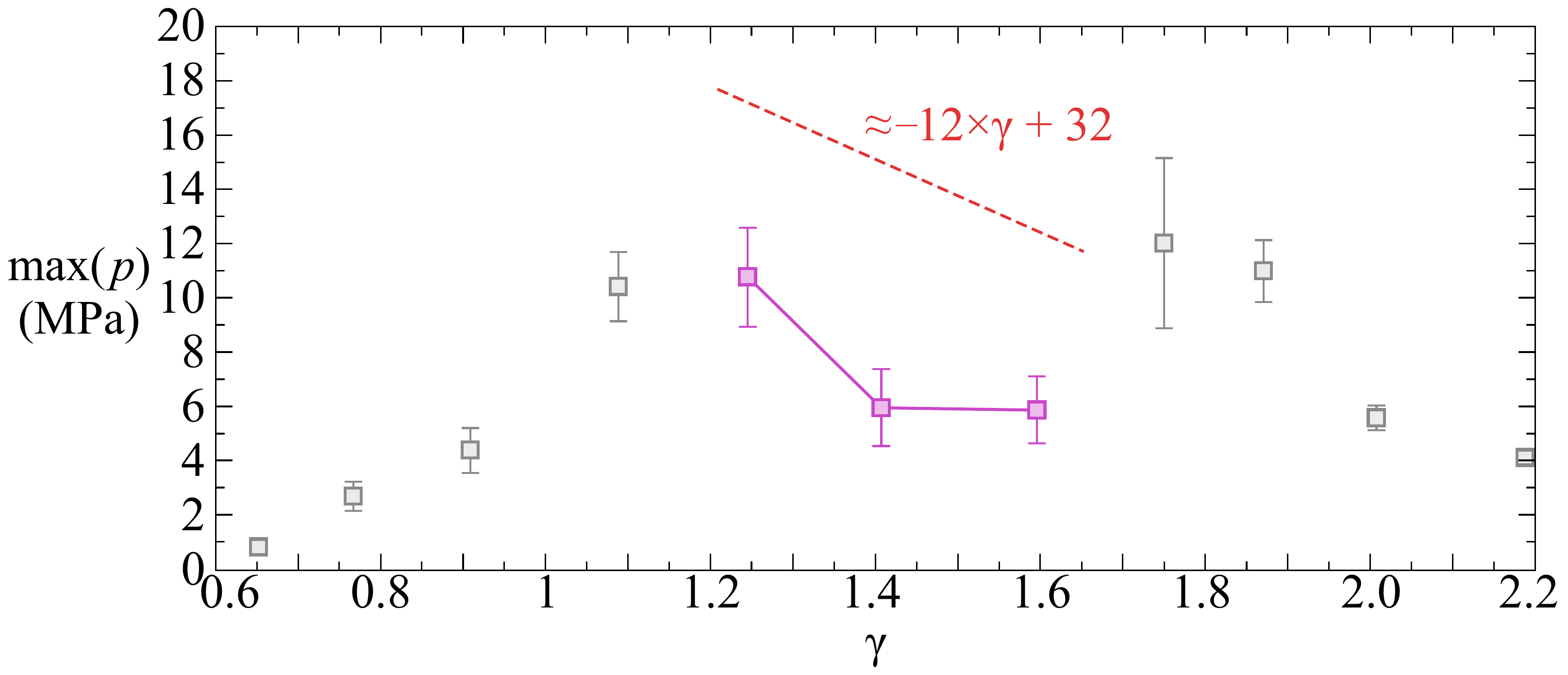}}
\caption{Second regime for $\gamma \in [1.24,\ 1.60]$: (a--c) Time evolution of wall pressure (left panels), microjet visualizations (top right row), shock wave visualizations (bottom right row. (d) Maximum pressure at the wall (markers) corresponding with linear (dashed).}
\label{fig:SecondCollapseMixedFirst}
\end{figure}

\clearpage

During this expansion, Raleigh-Taylor instability (RTI) can potentially develop, as discussed by \cite{Lindau2003} potentially causing the fragments. Recent studies highlight that the fragmentation observed is primarily due to Blake's splashing \citep{Splashing1, Splashing2}.  When a jet penetrates a bubble, it triggers splashing upon hitting the wall, causing radial flow and forming micro-bubble clouds in a toroidal shape. This splashing significantly impacts the interface of the rebounding bubble. Nevertheless, we speculate that the RTI will still be developed (see fig.\ref{fig:PTFandSTF}(a)), and be stronger when $\gamma\uparrow$ aiding the toroidal fragmentation along with the Blake's splashing, collapsing as STF, and PTF. While a weaker RTI only occurs as $\gamma$ decreases, the significant energy radiated during the first collapse results in a decrease in pressure as $\gamma$ decreases during the second collapse.

\begin{figure}
\centering
  \subfloat[$\gamma = 1.75$]{\vspace{-0.2cm}\includegraphics[height=0.25\textwidth]{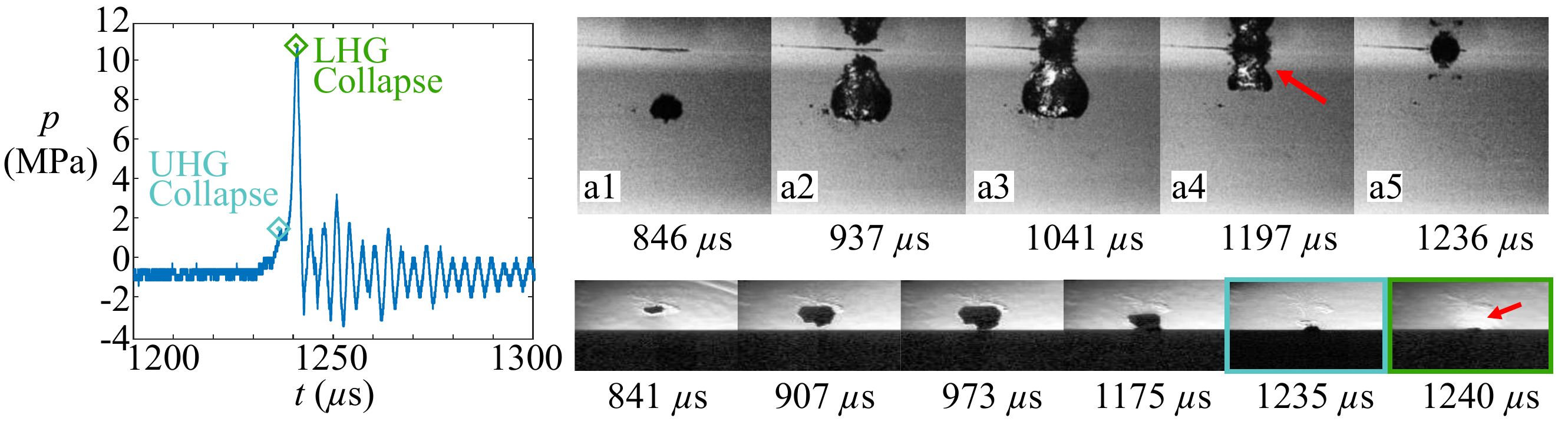}} \\
  \subfloat[$\gamma = 1.87$]{\vspace{-0.2cm}\includegraphics[height=0.25\textwidth]{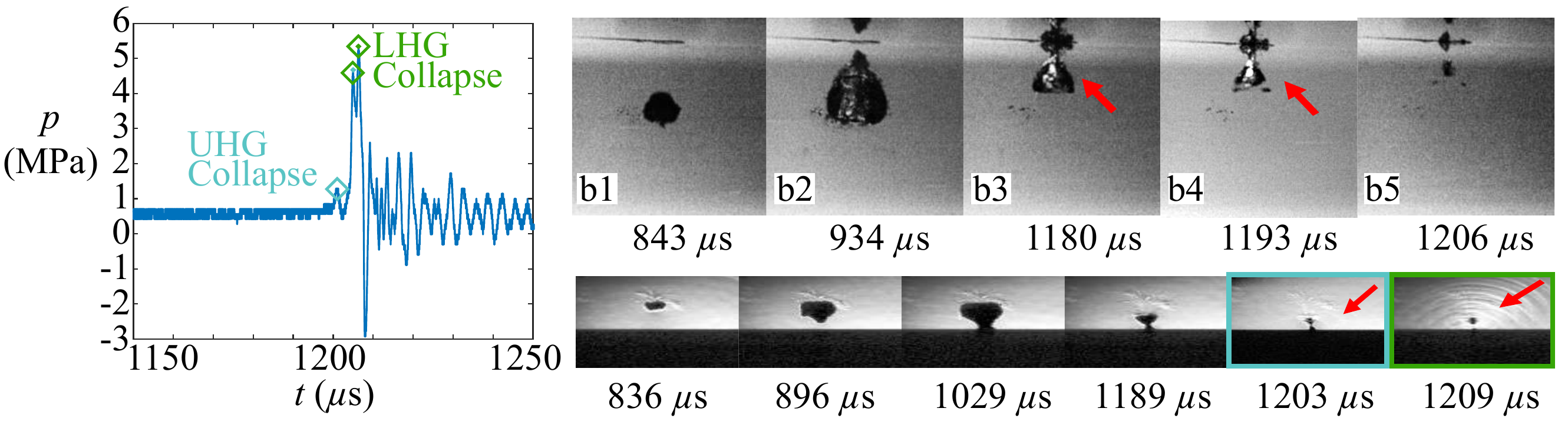}} \\
  \subfloat[$\gamma = 2.01$]{\vspace{-0.2cm}\includegraphics[height=0.25\textwidth]{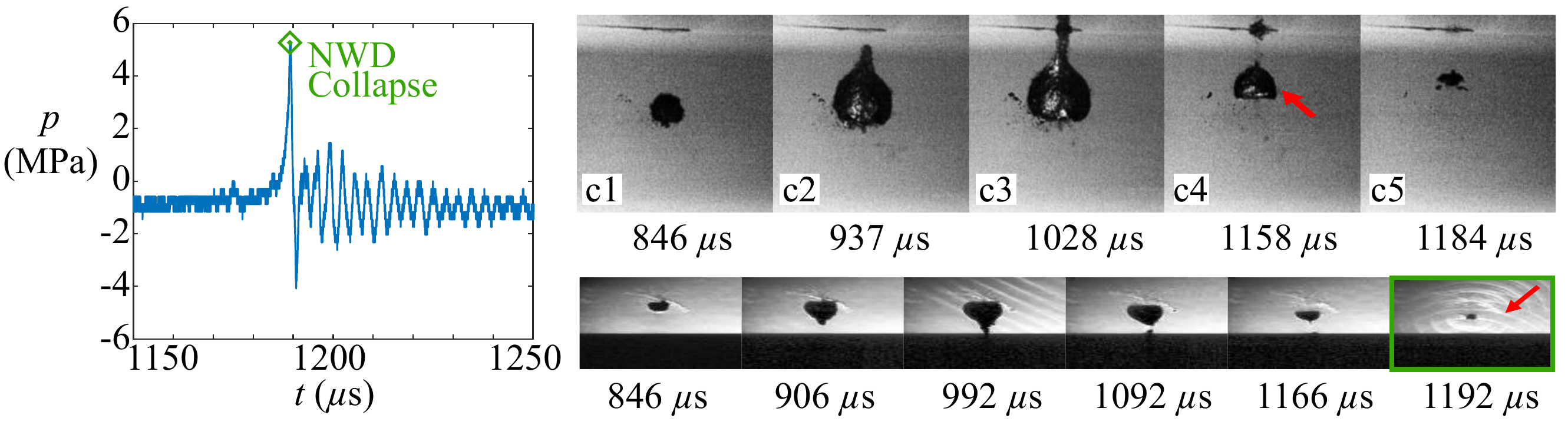}} \\
  \subfloat[$\gamma = 2.19$]{\vspace{-0.2cm}\includegraphics[height=0.25\textwidth]{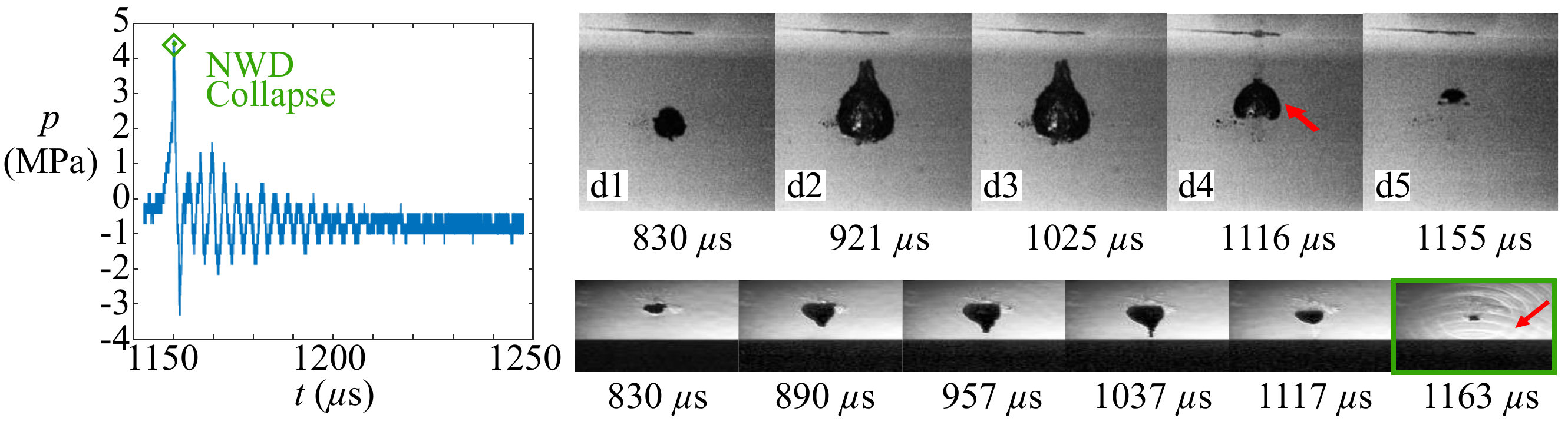}} \\
  \subfloat[maximum pressure at the wall during the second collapse]{\vspace{-0.2cm}\includegraphics[height=0.325\textwidth]{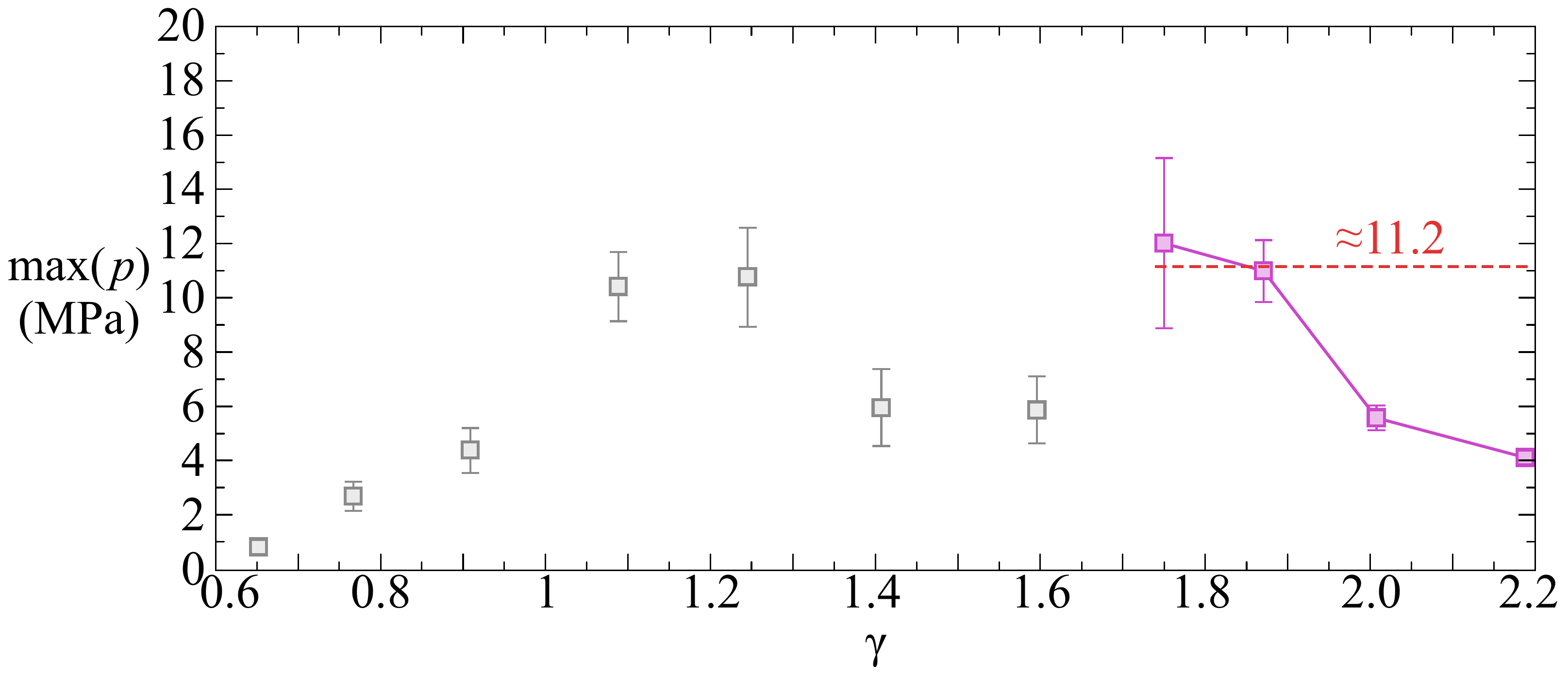}}
\caption{Third regime for $\gamma \in [1.75,\ 2.19]$: (a--d) Time evolution of wall pressure (left panels), microjet visualizations (top right row), shock wave visualizations (bottom right row). (e) Maximum pressure at the wall (markers) corresponding with constant fit (dashed).}
\label{fig:SecondCollapseTipFirst}
\end{figure}

 With a further increase in the stand-off ratio, we transition from the results shown in fig. \ref{fig:SecondCollapseTorusFirst}
 to those in fig. \ref{fig:SecondCollapseMixedFirst}. Similarly, fig. \ref{fig:SecondCollapseMixedFirst}(d) displays the pressures recorded during the second collapse in this second regime ($\gamma \in [1.24,\ 1.60]$), along with a linear fit derived from the first collapse data. In this regime, we still observe the collapse mechanism where the STF and PTF collapse sequentially. However, the toroidal bubble now dissociates into more fragments, leading to multiple STF collapses (depicted by light-blue markers in fig. \ref{fig:SecondCollapseMixedFirst}) and the emission of multiple shock waves. 
 
 This fragmentation may be attributed to the stronger development of the Raleigh-Taylor instability as $\gamma$ increases. Similar to the interaction observed between torus collapse and tip collapse, we observe strong interactions between STF collapses and PTF collapse, dissipating some energy through shock refractions before reaching the wall. This results in a trend similar to the negative slope linear pattern observed during the first collapse when $\gamma\uparrow$.

Finally, a third regime for the second collapse is also observed upon further increasing the stand-off ratio ($\gamma \in [1.75,\ 2.19]$). The corresponding pressures observed are shown in fig. \ref{fig:SecondCollapseTipFirst}.  Interestingly, similar to the absence of toroidal collapse during the first collapse due to delayed microjet formation, we also do not observe the toroidal bubble formation during the bubble rebound phase. This is likely due to the increased distance from the wall. 

We see two different mechanisms in this regime, and it is the formation of hour-glass bubble ($\gamma \in [1.75,\ 1.87]$), and the near-wall detached (NWD) bubble ($\gamma \in [2.01,\ 2.19]$). The NWD bubble collapse occurs when the near-wall interface separates the main bubble (see fig. \ref{fig:SecondCollapseTipFirst}(c4,d4)). The further collapse of the NWD bubble behaves quite similar to the spherical shock wave (see left panels of fig. \ref{fig:SecondCollapseTipFirst}(c,d)).

The formation of hour-glass bubbles prior to collapse can be observed in fig. \ref{fig:SecondCollapseTipFirst} (a4, b3, b4). These hour-glass bubbles collapse into two fragments: the upper hour-glass (UHG) bubble and the lower hour-glass (LHG) bubble. The UHG collapses first, followed by the LHG (see light-blue and green markers in the left panels of fig. \ref{fig:SecondCollapseTipFirst}). Corresponding shock emissions are visible in the images shown on the right. This collapse mechanism positions part of the rebounded bubble closer to the wall, thereby increasing the pressure. 

As a result, we observe a notable increase in the pressure trend between $\gamma=1.87$ to $2.01$, where the collapse mechanism changes. Although there is a slight variation in the mechanism between $\gamma=1.75$ and $1.87$, we still observe a similar spherical shock mechanism overall as explained during the first collapse in this regime.

\clearpage

\section{Conclusions}\label{sec:Conclusions}
 \vspace{5pt}
The dynamics of a single laser-induced cavitation bubble generated close to a rigid wall has been studied experimentally and will be investigated numerically in two follow-up studies. A pressure sensor mounted on the wall has been used to quantify the pressure magnitude experienced by the wall. Such a sensor has been used in combination with an hydrophone that served to measure the acoustic transients of the shock wave emitted by the bubble collapses. The corresponding pressure forces are recorded to understand and correlate them with one of the two mechanisms of cavitation erosion: microjet and shock-wave impacts on the rigid wall.   

The experimental evaluation of the microjet velocity and its corresponding impact time on the wall have been carried out by tracking the far- or near-wall bubble interface. They have been compared with the shock wave visualizations of the first and second bubble collapses in order to explain which of the two mechanisms (microjet inertia or shock wave) is responsible for the maximum wall pressure, hence for eventual wall erosion. Our high-speed visualizations and pressure sensor measurements let us conclude that the \textit{shock wave}, and not the microjet, is the dominant mechanism responsible for the sudden wall pressure increase observed in a single cavitation bubble collapsing near the rigid wall for all stand-off ratios considered, i.e. $\gamma \in [0.65,\ 2.19]$. 

Additionally, three bubble collapse regimes have been clearly identified for the first collapse: (i) a purely torus collapse regime for $\gamma \leq 1.09$, (ii) a mixed tip-and-torus collapse regime for $1.24 \leq \gamma \leq 1.60$, and (iii) a purely tip collapse regime for $1.75 \leq \gamma$. The dynamics of the second bubble collapse has been correspondingly characterized, demonstrating that second-collapse erosion significantly depends on the first collapse regime, hence on $\gamma$.

\bibliographystyle{jfm}
\bibliography{jfm-instructions}

\end{document}